\documentclass[review]{elsarticle}
\usepackage{amsmath}
\usepackage{amsfonts}
\usepackage{amssymb}
\usepackage{graphicx}
\usepackage{color}
\usepackage{bm}
\usepackage{float}
\usepackage{mathrsfs}
\usepackage{algpseudocode}
\usepackage{multicol}
\usepackage{hyperref}
\usepackage{algorithm}
\usepackage{algpseudocode}
\usepackage{smartdiagram}
\usepackage{tikz}
\usepackage{array}
\newcolumntype{C}[1]{>{\centering\arraybackslash}p{#1}}
\newcolumntype{R}[1]{>{\raggedleft\arraybackslash}p{#1}}
\newcolumntype{L}[1]{>{\raggedright\arraybackslash}p{#1}}
\usetikzlibrary{shapes.geometric, arrows}
\usepackage{subfig}
\usepackage[section]{placeins}

\newcommand{\abs}[1]{\left|#1\right|}

\newcommand{\norm}[1]{\left|\left|#1\right|\right|}
\newcommand{\cald}{\mathcal{D}}

\newcommand{\calF}{\mathcal{F}}
\newcommand{\bbe}{\mathbb{E}}

\newcommand{\bfPhi}{\mathbf{\Phi}}
\newcommand{\itPhi}{{\Phi}}
\newcommand{\bfu}{\mathbf{u}}
\newcommand{\bfv}{\mathbf{v}}
\newcommand{\bfx}{\mathbf{x}}

\newcommand{\bfy}{\mathbf{y}}
\newcommand{\bfY}{\mathbf{Y}}

\newcommand{\bfV}{\mathbf{V}}
\newcommand{\bfC}{\mathbf{C}}

\newcommand{\bft}{\mathbf{t}}
\newcommand{\bff}{\mathbf{f}}
\newcommand{\bfw}{\mathbf{w}}

\newcommand{\bbR}{\mathbb{R}}
\newcommand{\bbP}{\mathbb{P}}

\newcommand{\bsalpha}{\boldsymbol{\alpha}}
\newcommand{\bseta}{\boldsymbol{\eta}}

\newcommand{\bsxi}{\boldsymbol{\xi}}

\makeatletter
\def\ps@pprintTitle{%
 \let\@oddhead\@empty
 \let\@evenhead\@empty
 \def\@oddfoot{}%
 \let\@evenfoot\@oddfoot}
\makeatother

\usepackage{lineno,hyperref}
\modulolinenumbers[1]
\usepackage{float}
\floatstyle{plaintop}
\restylefloat{table}
\usepackage{array}
\newcolumntype{C}[1]{>{\centering\arraybackslash}p{#1}}
\newcolumntype{R}[1]{>{\raggedleft\arraybackslash}p{#1}}
\newcolumntype{L}[1]{>{\raggedright\arraybackslash}p{#1}}
\makeatletter
\def\@author#1{\g@addto@macro\elsauthors{\normalsize%
    \def\baselinestretch{1}%
    \upshape\authorsep#1\unskip\textsuperscript{%
      \ifx\@fnmark\@empty\else\unskip\sep\@fnmark\let\sep=,\fi
      \ifx\@corref\@empty\else\unskip\sep\@corref\let\sep=,\fi
      }%
    \def\authorsep{\unskip,\space}%
    \global\let\@fnmark\@empty
    \global\let\@corref\@empty  
    \global\let\sep\@empty}%
    \@eadauthor={#1}
}
\makeatother
\usepackage{tikz,bm,adjustbox}
\usepackage{fancyhdr}
\usetikzlibrary{arrows}
\tikzstyle{line} = [draw, -latex',line width=0.5mm,]
\tikzstyle{Rec} = [rectangle, draw, fill=white!100,
text width=10em, text centered,line width=0.5mm, rounded corners, minimum height=5em]
\usepackage{placeins}
\let\Oldsection\section
\renewcommand{\section}{\FloatBarrier\Oldsection}

\let\Oldsubsection\subsection
\renewcommand{\subsection}{\FloatBarrier\Oldsubsection}

\let\Oldsubsubsection\subsubsection
\renewcommand{\subsubsection}{\FloatBarrier\Oldsubsubsection}
\begin{document}

\title{High-dimensional Stochastic Inversion via Adjoint Models and Machine Learning}
\author{Charanraj A. Thimmisetty \fnref{fn1}}
\author{Wenju Zhao \fnref{fn3}}
\author{Xiao Chen\corref{cor1}\fnref{fn1}}
\ead{chen73@llnl.gov}
\author{Charles H. Tong \fnref{fn1} }
\author{Joshua A. White \fnref{fn2}}

\fntext[fn1] {Center for Applied Scientific Computing, Lawrence Livermore National Laboratory, Livermore, California, USA.}
\fntext[fn2] {Atmospheric, Earth and Energy Division, Lawrence Livermore National Laboratory, Livermore, California, USA.}
\fntext[fn3] {Department of Scientific Computing, Florida State University, Tallahassee, FL USA.
\vspace{0.2in}
\begin{flushright}
\tiny{LLNL-JRNL-744399-DRAFT}
\end{flushright}
}

\cortext[cor1]{Corresponding author}

\begin{abstract}
Performing stochastic inversion on a computationally expensive forward simulation model with a high-dimensional uncertain parameter space (e.g. a spatial random field) is computationally prohibitive even with gradient information provided. Moreover, the `nonlinear' mapping from parameters to observables generally gives rise to  non-Gaussian posteriors even with Gaussian priors, thus hampering the use of efficient inversion algorithms designed for models with Gaussian assumptions.  In this paper, we propose a novel Bayesian stochastic inversion methodology, characterized by a tight coupling between a gradient-based Langevin Markov Chain Monte Carlo (LMCMC) method and a kernel principal component analysis (KPCA). This approach addresses the `curse-of-dimensionality'  via KPCA to identify a low-dimensional feature space within the high-dimensional and nonlinearly correlated spatial random field. Moreover, non-Gaussian full posterior probability distribution functions are estimated via an efficient LMCMC method on both the projected low-dimensional feature space and the recovered high-dimensional parameter space. We demonstrate this computational framework by integrating and adapting recent developments such as data-driven statistics-on-manifolds constructions and reduction-through-projection techniques to solve inverse problems in linear elasticity.
\end{abstract}


\maketitle	

\section{Introduction}
Computational science and engineering have enabled researchers to model complex physical processes in many disciplines---e.g. mechanical behavior~\cite{anandarajah2011computational}, climate projection~\cite{kirtman2013near}, subsurface flow and reactive transport~\cite{dagan2005subsurface}, seismic wave propagation~\cite{kennett2013seismic, graves1996simulating}, and power grid planning~\cite{kundur1994power}. However, uncertainty in the model parameters makes the underlying problems essentially stochastic in nature. Applying uncertainty quantification (UQ) to improve model predictability usually requires solving an inverse problem (inverse UQ) by `fusing' prior knowledge, simulations, and experimental observations. Deterministic approaches to solve inverse problems, such as regularized weighted nonlinear least squares methods, are capable of providing an optimal statistical estimator with associated error bars for the inverse solutions. However, these approaches, by their deterministic nature, cannot produce solutions with a full description of the posterior probability density functions (pdf). Unlike deterministic inversion, stochastic inversion aims to provide this fuller description. A pdf representation  is critical for prediction of system performance, so that appropriate decisions can be made according to the probability and risk associated with specific events.

Bayesian inference provides a systematic framework for integrating prior knowledge and measurement uncertainties to compute detailed posteriors~\cite{tarantola2005inverse}. However, it can be computationally intractable~\cite{martin2012stochastic} to compute the full pdf for parameters assigned to each grid point of a discretized parametric random field---i.e., the curse of dimensionality~\cite{martin2012stochastic}. Moreover, unreasonable choices of prior knowledge due to ignorance of the information embedded in the underlying dataset for model parameters can have major effects on inferring the posterior pdf. In addition, the nonlinear mapping between the observables and parameters leads to non-Gaussian posteriors even with additive noise and Gaussian prior assumptions~\cite{martin2012stochastic}. In general, it is mathematically challenging to sample directly from a non-Gaussian and multi-modal posteriors  especially in a very high-dimensional random space. MCMC methods are relevant techniques for sampling non-standard posteriors. Despite the computational intensity encountered in MCMC, these methods have grown in rigor and sophistication with recent technical developments such as delayed rejection (DR)~\cite{green2001delayed, mira2001ordering}, adaptive Metropolis (AM)~\cite{haario1999adaptive, haario2001adaptive,tierney1999some}, delayed rejection adaptive Metropolis (DRAM)~\cite{roberts1998optimal},  stochastic Newton~\cite{martin2012stochastic}, Langevin~\cite{haario2006dram} and transport map accelerated MCMC~\cite{parno2014transport}.

The gradient-free MCMC methods, e.g., random walk MCMC, DR, AM, and DRAM, become computationally intractable as the size of the parameter space increases just moderately. Even though the gradient-enhanced MCMC algorithms such as Langevin ~\cite{haario2006dram} and stochastic Newton methods ~\cite{martin2012stochastic} have decreased the computational complexity of classical MCMC   to $O(n^{1/3})$, expensive high-fidelity forward models,  mesh-defined high-dimensional parameter spaces, and multi-modal non-Gaussianity cause significant computational challenges in practice, rendering these algorithms unsuitable for large-scale, real-world problems.

One way to address the computational complexity of MCMC is through a construction of low-fidelity surrogate models using design of experiments (DOE) with the help of machine learning techniques, e.g., global polynomials~\cite{ghanem2003stochastic,marzouk2009stochastic,marzouk2007stochastic}, radial basis functions~\cite{bliznyuk2012local, joseph2012bayesian}, Gaussian processes~\cite{rasmussen2006gaussian}, neural networks~\cite{funahashi1989approximate,hornik1989multilayer}, and/or proper orthogonal decomposition (POD) based reduced modeling. The use of low-fidelity models, based on surrogate and/or reduced-order modeling, greatly reduces the computational cost of the stochastic inversion. Low-fidelity model-based stochastic inversion, however, tends to produce entirely different inverse solutions or sub-optimal solutions compared to the true posterior obtained by the corresponding high-fidelity model-based stochastic inversion.

Instead of performing forward model reduction, another way to reduce MCMC complexity is through control reduction, by performing Bayesian inference in a low-dimensional subspace embedded in the high-dimensional parameter space, while still using the high-fidelity forward model constrained onto this low-dimensional space. Karhunen-Lo\'eve or principal component analysis (PCA) is a well-known choice for such parametric control dimension reduction. Traditionally, PCA is designed for the representation of linear correlation of the underlying data. Many realistic parametric random fields, however, exhibit non-linear correlations in the underlying data. The subspace spanned by PCA might not even cover the solution domain. Furthermore, one has to perform an exhaustive search to reach to the true posterior due to the widely scattered reduced space represented by the linear PCA-extracted subspace.

The method proposed here uses unsupervised learning techniques to obtain relevant subspaces. Recent advances in unsupervised machine learning algorithms have provided ways to explore non-linear datasets using manifold learning techniques. Specifically, kernel PCA~\cite{scholkopf1997kernel} (KPCA) has been demonstrated to perform better clustering than linear PCA on complex non-linear data. Recently, Sarma~\cite{Sarma2008} and Ma~\cite{Ma:2011:KPC:2016171.2016423} demonstrated the efficiency and benefits of KPCA for deterministic forward and inverse uncertainty propagation.

Here, we propose a novel framework for efficient stochastic inversion using adjoint partial differential equations (PDEs), automatic differentiation (AD), and KPCA. We demonstrate our approach on a stochastic linear elasticity inversion problem.  For this application, a full statistical analysis in the high-dimensional ``ambient" space spanned by grid-defined model parameters is computationally prohibitive. In addition, the model output is a high-dimensional vector space defining the solution variables over the whole spatial discretization. Thus, we have the challenge of an ambient space where  each measurement is a high-dimensional vector obtained as an expensive model evaluation. The solution, however, is constrained: it does not occupy the whole ambient space, but merely a low-dimensional manifold within it. Because only a low-dimensional probability space needs to be explored, we can design novel algorithms to accelerate the convergence of MCMC algorithms.

We use the following sequence to reduce the computational burden of solving large-scale  stochastic inverse problems in elasticity.  The methods studied here are general, however, and can be extended to many other application areas.
\begin{itemize}
\item The linear elastic model is described by a system of  self-adjoint PDEs that facilitate computation of the cost functional gradient with respect to the high dimensional, grid-defined model parameters.  At any configuration, the gradient of the cost functional with respect to the model parameters may be computed using two simulations (a forward and adjoint simulation).
\item Using geostatistical methods---specifically the single normal equation simulation (SNESIM) algorithm~\cite{strebelle2002conditional}---we generate statistical realizations of a complex property model used as the basis for prior knowledge. Then, a low-dimensional feature space is obtained by performing KPCA on the generated geostatistical realizations.
\item The feature random variables obtained from the KPCA are uncorrelated but not Gaussian. In general, Bayesian frameworks requires frequent sampling on these feature random variables. To improve sampling efficiency, we sample them using a polynomial chaos expansion (PCE) coupled with an inverse cumulative distribution function (ICDF) transformation.
\item We then construct an automatic differentiation-based discretized adjoint model of the KPCA-based and PCE-based ICDF transformation, and couple the discretized adjoint model with the high-fidelity adjoint PDE model.  This approach provides gradients of the cost functional with respect to the low-dimensional feature random variables.
\item Bayesian inference is then performed on the low-dimensional feature space using an efficient LMCMC scheme. The convergence rate of this KPCA and gradient-based stochastic inversion through MCMC is greatly improved, thanks to the nonlinear control reduction with good classification and clustering properties.
\item Unlike traditional machine learning problems, this process in each MCMC iteration step requires the projection of the low-dimensional feature space back to the high-dimensional parameter space, since the high-fidelity forward models are functions of grid-defined model parameters.  The projection is obtained by exploring both local fixed-point iteration and non-iterative algebra approaches.
\item This projection from the feature space back to parameter space gives us access to posterior pdf of the grid-defined high-dimensional model parameters.
\end{itemize}

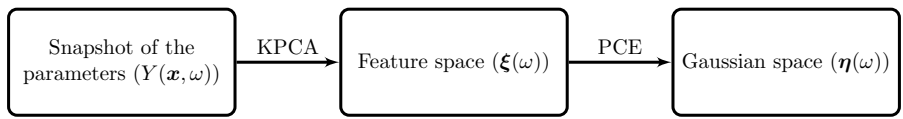
\begin{figure}[p]
\centering
 \begin{adjustbox}{max width=0.98\textwidth}
\begin{tikzpicture}[node distance = 5.5cm, auto]
\node [Rec] (B1){Snapshot of the parameters ($Y(\bm x,\omega)$) };
\node [Rec,right of=B1] (B2) {Feature space ($ {\color{black}\bm \xi}(\omega)$)};
\node [Rec,right of=B2] (B3) {Gaussian space $( { \color{black}\bm \eta} (\omega))$};
\path [line] (B1) -- node [midway] {KPCA}(B2);
\path [line] (B2) -- node [midway] {PCE}(B3);
 \end{tikzpicture}
 \end{adjustbox}
 \caption{Mapping from the parameter space to Gaussian space}\label{fig:alg1}
  \end{figure}

\begin{figure}[p]
 \centering
 \begin{adjustbox}{max width=0.98\textwidth}
\begin{tikzpicture}[node distance = 6.2cm, auto]
\node [Rec] (B4) {Sample of  ${\color{black}\bm\eta(\omega)}$};
\node [Rec,right of=B4] (B5) {Sample of  ${\color{black}\bm\xi}(\omega)$};
\node [Rec,right of=B5] (B6) {Sample of  $Y(\omega)$};
\node [Rec,below of=B6,yshift=7em] (B7) {Forward model};
\node [Rec,left of=B7] (B8) {Adjoint model};
\node [Rec,above of=B8,yshift=-12em,draw=none,minimum height=0.1em] (B8_1) {Measurements};
\node [Rec,left of=B8] (B9) {Cost function $(J)$,  gradient of the cost function ${\color{black}(\frac{\partial J}{\partial \bm\eta})}$};

\path [line] (B4) -- node [midway] {PCE}(B5);
\path [line] (B5) -- node [midway] {Pre-imaging}(B6);
\path [line] (B6) -- node [midway,text width=8em] {Model \\parameters $(\lambda,\mu)$}(B7);
\path [line] (B7) -- node [midway] {Predictions}(B8);
\path [line] (B8) -- node [midway] {AD adjoint}(B9);
\path [line] (B9) -- node [midway] {LMCMC}(B4);
\path [line] (B8_1) --(B8);
 \end{tikzpicture}
\end{adjustbox}
\caption {Posterior sample generation with proposed approach}\label{fig:alg2}
 \end{figure}
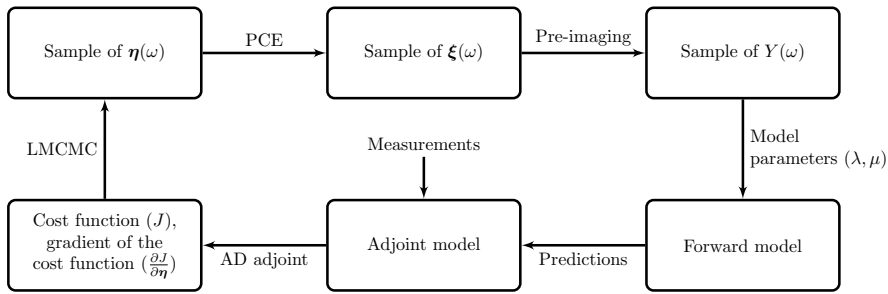

The remainder of this paper is organized as the following. Section~\ref{sec:mf} provides the mathematical framework of our procedure, providing a detailed derivation of each step in the proposed method. To help guide the reader through these developments, Table~\ref{tab1} provides a summary of the proposed workflow and the key challenges each step seeks to address; Figures~\ref{fig:alg1} and \ref{fig:alg2} provide the flowchart of the mapping from the parameter space to Gaussian space and posterior sample generation with proposed approach, respectively; and Algorithms 1 and 2 provide a concise summary of the steps necessary to implement the methodology.  In Section~\ref{sec:ns}, we apply this methodology to identify elastic properties of a geologically complex system. Section~\ref{sec:ds} gives some insights on the advantages of KPCA and the implementation of the proposed method for stochastic inversion. Finally, conclusions are given in Section~\ref{sec:co}, with an outline of future work.

\begin{table}[p]
\centering
\footnotesize
\caption{Summary of the proposed approach}\label{tab1}
\begin{tabular}{|L{0.08\linewidth}L{0.20\linewidth}L{0.15\linewidth}L{0.45\linewidth}|}
\hline
Section& Challenge& Approach& Explanation\\
\hline
2.1& High-fidelity gradient computation & Adjoint gradient & Numerical gradient computation using finite difference methods requires many forward model runs. Here, the self-adjoint PDE allows us to compute gradients in the parameter space with two model runs (a forward and adjoint simulation). \\ \hline
2.2& High dimensionality of the parameters& KPCA& KPCA is used to find a low-dimensional feature space where the solution is not an outlier in the prior probability space.                                                            \\ \hline
2.3  & Sampling non-Gaussian feature random variables & PCE & KPCA feature random variables are uncorrelated but dependent on non-Gaussian random variables.
 An ICDF transformation is used to build the PCE of the feature random variables to facilitate efficient sampling.                                                                       \\ \hline
2.4 & Ill-posedness of the inverse problem                           & Bayesian inference               & Sparse and noisy measurements and high-dimensionality of the parameter space make the inverse problem ill-posed. Bayesian inference provides a systematic way to address these problems and provides a probabilistic inverse solution.                                                                                      \\ \hline
2.5--2.6 & Computational intractability of the MCMC                       & LMCMC and automatic differentiation & Gradient free MCMC ($O(n)$) quickly runs into computational intractability as the problem size of the parameter space increases. Gradient based LMCMC $(O(n^{1/3}))$ is used to make the solution tractable by performing inversion in the lower-dimensional feature space and leveraging the derivative information of feature random variables obtained by automatic differentiation.\\
\hline
\end{tabular}
\end{table}

\section{Formulation}\label{sec:mf}
\subsection{Elasticity model}
\label{elasticitymodel}
This section introduces a model problem to test the proposed inversion approach: the deformation behavior of a linear elastic body under mechanical loads. The governing PDE is a linear momentum balance equation involving two elastic coefficients, the Lam\'e parameters of the material \cite{marsden:1983}. The goal is to estimate these material properties based on sparse measurement data and prior knowledge.

Let  the physical domain $\cald \subset \bbR^d,\: d = 2,\: 3$, be a bounded, connected, open  and Lipschitz
continuous domain with a boundary $\Gamma = \partial \cald$.
Assume $\Gamma_D$ and $\Gamma_N$ are two subsets of $\Gamma$ such that $\Gamma_D \cap \Gamma_N = \emptyset$
and $\Gamma_D \cup  \Gamma_N = \Gamma$. Let Dirichlet and Neumann  boundary conditions (prescribed displacements and prescribed tractions) be
specified along $\Gamma_D$ and $\Gamma_N$, respectively.  For an  integer $m\geq 0$, we follow
the classical notation of a  standard Sobolev space  $H^m(\cald)$ with norm $\norm{\cdot}_m$ in
accordance with Adams et al. \cite{MR2424078}.

To express the governing PDE in variational form---suitable for finite element discretization---let
\begin{align}
\mathcal{U} &= \{ \bfu: \cald \to \bbR^d \; | \; \bfu \in H^1, \bfu = \overline{\bfu} \; \text{on} \; \Gamma_D \} ,\\
\mathcal{V} &= \{ \bfv: \cald \to \bbR^d \; | \; \bfv \in H^1, \bfv = \mathbf{0} \; \text{on} \; \Gamma_D \}
\end{align}
be spaces of trial displacement fields $\bfu(\bfx)$ and weighting functions $\bfv(\bfx)$. Prescribed displacement boundary conditions $\overline{\bfu}$ are assigned on $\Gamma_D$. The weak problem is then to find $\bfu \in \mathcal{U}$ such that, for all $\bfv \in \mathcal{V}$, the following linear momentum balance equation is satisfied,
\begin{equation}
a(\bfv,\bfu) = (\bfv,\bff) + (\bfv,\bft)_{\Gamma_N}{\color{black},}
\label{eq:model}
\end{equation}
where the respective bilinear forms are
\begin{align}
a(\bfv,\bfu) &= \int_\cald \lambda (\nabla \cdot \bfv ) (\nabla \cdot \bfu) \,d\cald + \int_\cald 2\mu (\nabla^s \bfv : \nabla^s \bfu) \,d \cald {\color{black},}\\
(\bfv,\bff) &=
\int_\cald \bfv \cdot \bff \,d\cald  {\color{black},}\\
(\bfv,\bft)_{\Gamma_N} &=\int_{\Gamma_N} \bfv \cdot \bft \, d \Gamma {\color{black}.}
\end{align}

Here, $\nabla^s = (\nabla + \nabla^T) / 2 $ is the symmetric gradient operator, $\bff$ is a body force due to self-weight, and {\color{black}$\bft$  is an externally applied traction on $\Gamma_N$. The two material coefficients $\lambda(\bfx)$ and $\mu(\bfx)$ are the Lam\'e parameters describing the elastic properties of the body.

For brevity, we omit most of the details of the finite element discretization, as they are standard \cite{hughes:2000}.  We introduce a partition of $\cald$ into non-overlapping elements $\cald^e$.  On this mesh, both vector and scalar fields are discretized using {\color{black} bilinear or trilinear basis functions }$\{\phi^a\}$ as
\begin{align}
\bfu^h(\bfx) &= \sum_{a=1}^{n_\text{nodes}}  \bfu^a \phi^a (\bfx) {\color{black},} \\
\bfv^h(\bfx) &= \sum_{a=1}^{n_\text{nodes}}  \bfv^a \phi^a (\bfx) {\color{black},} \\
\lambda^h(\bfx) &= \sum_{a=1}^{n_\text{nodes}}  \lambda^a \phi^a (\bfx) {\color{black},} \\
\mu^h(\bfx) &= \sum_{a=1}^{n_\text{nodes}} \mu^a \phi^a (\bfx){\color{black},}
\end{align}
where the coefficients represent the nodal values of each field.  Introducing these discrete fields into the variation form (\ref{eq:model}), the problem can be recast as a discrete linear system
\begin{equation}
Au= {\color{black}b}
\label{eq:matrix}
\end{equation}
whose solution $u$ is an algebraic vector of unknown displacement components at the mesh nodes.  We will refer to the solution of this linear system as the \emph{forward simulation}.

The matrix $A$ depends on the material properties {\color{black}$\lambda^h(\bfx)$ and $\mu^h(\bfx)$. These material properties are assigned at each node of the mesh. Let $p$ denote an algebraic vector containing the property coefficients $\{\lambda^a,\mu^a \}$. The vector $p$ has {\color{black}dimension $2 \times n_\text{nodes}$.  The vector space of possible $p$ configurations is therefore extremely large for highly-refined meshes.  Attempting to solve an inverse problem for $p$ in this space is challenging. It will be even more challenging to provide the uncertainty information in this space.

Assuming discrete observations $u^\text{obs}$ are available in certain locations, a simple cost functional can be defined as
\begin{equation}
J(p) = \frac{1}{2} e^T D e \qquad \text{with} \qquad e_i = u_i  - u_i^\text{obs} { \color{black},  }
\end{equation}
where $D$ is a diagonal matrix containing weighting coefficients for each observation.  For a displacement component $u_i$ where no observational data is available, the corresponding diagonal entry $D_{ii}$ is zero.  Note that additional terms can be added to the cost functional to include regularization terms and other types of observational data beyond displacements.

The minimization of the cost functional is an optimization problem that can benefit from the calculation of gradient information.  In particular, the gradient vector $g$ has components
\begin{equation}
g_i = \frac{\partial J}{\partial p_i} = \frac{\partial J}{\partial u_j} \frac{\partial u_j}{\partial p_i} = e_k D_{kj} \frac{\partial u_j}{\partial p_i} {\color{black}.}
\end{equation}
Here, summation over repeated indices is implied.  By differentiating equation~(\ref{eq:matrix}) with respect to $p$, we find \cite{oberai2003solution,Oberai:2004}
\begin{equation}
\frac{\partial A_{mn}}{\partial p_i} u_n + A_{mj} \frac{\partial u_j}{\partial p_i} = 0 {\color{black}.}
\end{equation}
and therefore,
\begin{equation}
\frac{\partial u_j}{\partial p_i} = - A_{jm}^{-1} \frac{\partial A_{mn}}{\partial p_i} u_n {\color{black}.}
\end{equation}
Inserting this expression into the gradient formula and using the symmetry properties of $A$, the gradient can be expressed as
\begin{equation}
g_i = - w_m \frac{\partial A_{mn}}{\partial p_i} u_n {\color{black},}
\label{eqn:gradone}
\end{equation}
where the vector $w$ is the solution of the linear system,
\begin{equation}
A w = D e {\color{black}.}
\end{equation}
Note that this system is similar to equation (\ref{eq:matrix}) due to the self-adjoint nature of the underlying PDE.  We will refer to the solution of this system as the \emph{adjoint simulation}.  Once the fields {\color{black}$\bm u(\bfx)$ and $\bm w(\bfx)$} are computed by solving the forward and adjoint systems, equation~(\ref{eqn:gradone}) allows individual components of the gradient vector to be computed explicitly as
\begin{align}
g^{\mu}_i & = \frac{\partial J}{\partial \mu^i} = \int_{\cald} 2 \phi^i (\nabla^s \bfw^h : \nabla^s \bfu^h) d \cald {\color{black},} \\
g^{\lambda}_i & = \frac{\partial J}{\partial \lambda^i} = \int_{\cald} \phi^i (\nabla \cdot \bfw^h) (\nabla \cdot \bfu^h) d \cald {\color{black}.}
\end{align}


\subsection{Discretization of the random field and kernel principal component analysis}
The high dimensionality of the discretized parameter space can lead to intractability of the {\color{black} stochastic }  inversion problem. This section introduces a KPCA method to find a low-dimensional but relevant feature space.

To describe the stochastic nature of the PDE, let
 $\Omega$ be a sample space associated with probability triplet $(\Omega, \calF,  \bbP)$
where $\calF \subset 2^\Omega$ is a $\sigma$-algebra
of the events in $\Omega$ and $\bbP$ is the probability measure $ \bbP : \calF \rightarrow [0,1]$. We assume the two material coefficients $\mu( \bm x,\omega) : \cald \times
\: \Omega \rightarrow \bbR $ and  $\lambda( \bm x,\omega) : \cald \times \:\Omega \rightarrow \bbR $---the elastic Lam\'e parameters---are now random fields belonging to an infinite-dimensional probability space.

Let $Y(\bm x,\omega):=\text{ln}(\mu (\bm x, \omega))$ be a random field. The covariance function  can be defined as $C_Y(\bm x,\bm y)=<\tilde Y(\bm x,\omega) \tilde Y(\bm y,\omega)>_\omega$, where $\tilde Y(\bm x,\omega):= Y(\bm x,\omega) - <Y(\bm x,\omega)>_{\omega}$  and {\color{black}$< . >_{\omega}$ is an expectation operator. Assuming $C_Y$ is bounded, symmetric and positive definite, it can be represented as~\cite{courant1966methods}
\begin{equation}
C_Y(\bm x, \bm y)=\sum\limits_{i=1}^\infty  \gamma_i e_i(\bm x) e_i(\bm y),
\end{equation}
where  $\gamma_1\geq\gamma_2\geq\cdots > 0$ are  the eigenvalues, and
 $e_i(\bm x)$ and $e_j(\bm y)$  are deterministic and mutually orthogonal  functions,
\begin{equation}
\int_{\color{black}\cald} e_i(\bm x) e_j(\bm x) \: d\bm x=\delta_{ij}, \quad i,j\ge 1.
\end{equation}
Using Karhunen-Lo\`eve (KL) expansion, the random process $\overline Y(\bm x, \omega)$ can be expressed in terms of $e_i(\bm x)$ as
\begin{equation}
\overline Y(\bm x, \omega)=\sum\limits_{i=1}^\infty {\color{black}\xi_i}(\omega) \sqrt{\gamma_i} e_i(\bm x),\label{kl:inf}
\end{equation}
where {\color{black} $\{\xi_i(\omega)\} $}  are zero-mean and uncorrelated  random variables, i.e.,{\color{black} $<\xi_i(\omega)>_\omega=0$ and $<\xi_i(\omega) \xi_j(\omega)>_\omega=\delta_{ij}$}.
The eigenvalues \{$\gamma_i$\} and the eigenfunctions {\color{black}$ f_i(\bm x)$}  are obtained by solving the following integral equation either analytically or numerically,
{\color{black}
\begin{equation}
\int_{\cald} C_Y(\bm x, \bm y) f_i(\bm x) \: d\bm x=\gamma_i e_i(\bm y),\; i = 1,2, \dots. \label{KL:al}
\end{equation}
}
The attenuation of the eigenvalues \{$\gamma_i$\} allows truncation of the infinite sum in  Equation~\eqref{kl:inf}  up to $N_R$ terms,
\begin{equation}
\overline Y(\bm x, \omega)\approx \sum\limits_{i=1}^{N_R} \xi_i(\omega) \sqrt{\gamma_i} e_i(\bm x),
\end{equation}
where $N_R$ is the stochastic dimension. The  KL expansion is optimal~\cite{ghanem2003stochastic} in the sense that it
minimizes the mean-square error out of all possible orthonormal bases in $L^2(\cald \times\Omega)$.

In practice, a closed form expression for the $C_{Y}$ is rarely available. Instead, a numerical approximation to the $C_{Y} (\bm x, \bm y)$ is obtained using realizations of  $Y(\bm x , \omega)$ as:
\begin{equation}
 C_{Y} (\bm x, \bm y)\approx \frac{1}{M} \sum_{i=1}^{M} (Y(\bm x,\omega_i) - <Y(\bm x,\omega_i)>_{\omega}) (Y(\bm y,\omega_i) - <Y(\bm y,\omega)>_{\omega})^T,
\end{equation}
where $M$ is the number of realizations extracted from the random field $Y(\bm x,\omega)$. Given $C_{Y}$, approximation to Equation \eqref{KL:al} can be obtained using  the Nystrom algorithm~\cite{press2007numerical} as
\begin{equation}
\sum_{i=1}^{M} w_i  C_{Y} (\bm x_i, \bm y) e(\bm x_i)= \gamma e(\bm y). \label{eq:pca}
\end{equation}

Here, $M$  is the number of sample points where realizations ${\bm x}_i$'s are provided, and $w_i$'s are weights of  the quadrature rule.  Assuming we have enough sample points and equal weights $w_i=\frac{1}{M}$, equation (\ref{eq:pca}) can be solved by simple eigen-decomposition of $ C_{Y} (\bm x_i, \bm y)$, for which principal component analysis (PCA)~\cite{bishop2006pattern}  can be used to reduce the dimension.

The current data assimilation framework has the ability to infuse various sources of information into the Bayesian framework. For instance, the application considered in this paper is the elastic deformation of subsurface geologic formations under mechanical loads.  Along with displacement measurements (model solutions), we often have access to elasticity parameter measurements (hard data) at a few sparse locations obtained from wells. In addition, geophysical parameters can be obtained with 3D seismic observations (soft data). The soft and hard data are generally used to generate geostatistical realizations of model parameters. For instance, a simple geostatistical spatial random process for the prior parameter field can be obtained with two point statistical methods such as Kriging~\cite{cressie1990origins,isaaks1989applied,matheron1963principles}. A more general category of data-driven methods that build on soft and hard data measurements includes multi-point statistics (MPS)~\citep{strebelle2002conditional}, soft computing methods such as neural network, fuzzy logic, support vector machines~\citep{lim2003reservoir,nikravesh2001past,nikravesh2003soft,ouenes2000practical,gholami2012prediction} and Gaussian process on manifolds~\cite{Thimmisetty2017}. In the numerical examples, we will use MPS to generate elastic property models describing complex channelized structures frequently encountered in the subsurface.

The stochastic dimension of the prior model obtained using MPS is proportional to the number of finite element grid points in the simulation model. Equation~\eqref{eq:pca}, which is  equivalent of performing PCA of the covariance matrix, can be used to  reduce this dimension size. However, in general, PCA can only obtain efficient embeddings for linearly correlated data points. Recently,  Sarma~\cite{Sarma2008} and Ma~\cite{Ma:2011:KPC:2016171.2016423} have shown that KPCA is an appealing alternative for dealing with complex prior models.

We use two simple examples to demonstrate the desirable properties of KPCA. Figures ~\ref{fig:kpca-ex} (a) and (b) depict a classification problem where the
objective is to classify a XOR dataset~\cite{bishop2006pattern}. KPCA with a second-order polynomial
kernel can classify data perfectly, while  PCA  has a lower accuracy.
Figures ~\ref{fig:kpca-ex} (c) and (d) show another example~\cite{bishop2006pattern}, the goal of which is
to reduce the dimensionality of a non-linear dataset that lies across a curve. It indicates that a KPCA-based one-dimensional (1D) subspace is closer to true data than a PCA-based 1D subspace. In the following, we take advantage of both dimension reduction and improved feature representation properties of KPCA to increase the efficiency of stochastic inversion. Specifically, KPCA is used to find a low-dimensional and relevant feature space where the solution is not an outlier in the prior probability space.
\begin{figure}[h]
\centering
\subfloat[][]{\includegraphics[width=0.85\textwidth]{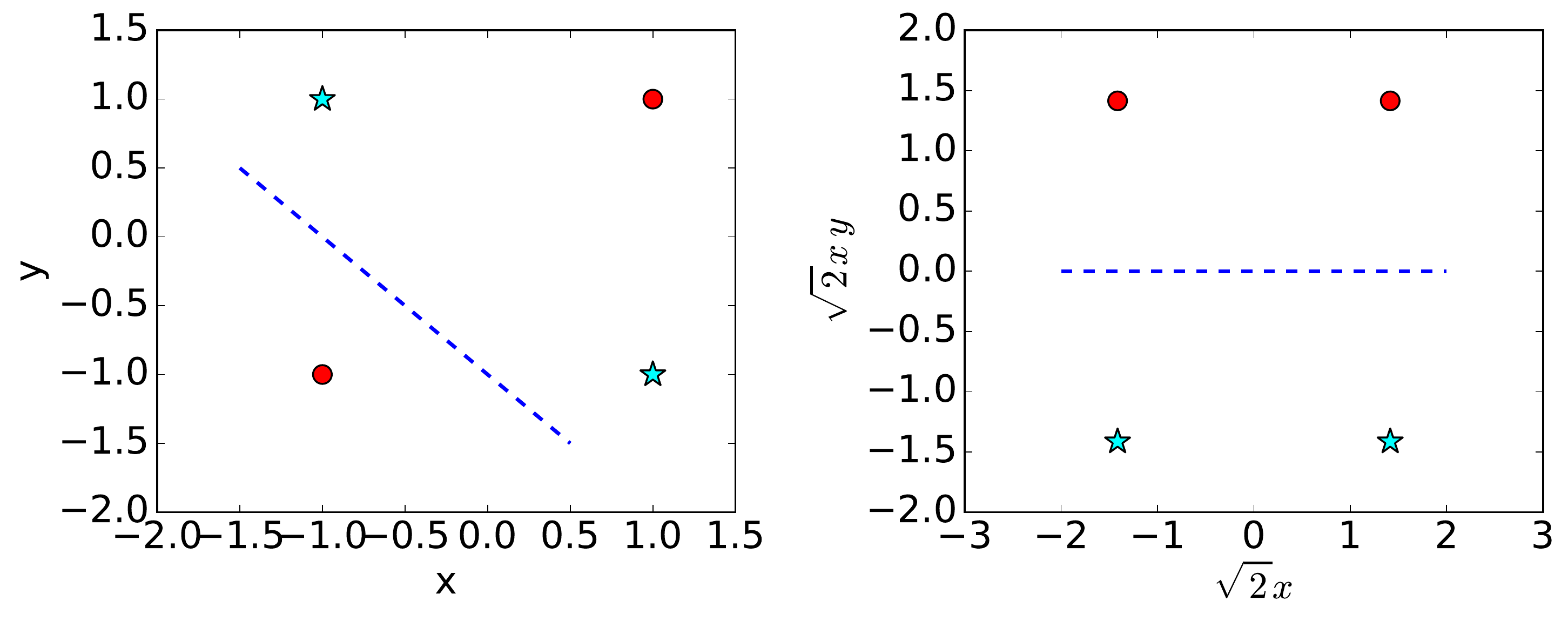}}\\
\subfloat[][]{\includegraphics[width=0.75\textwidth]{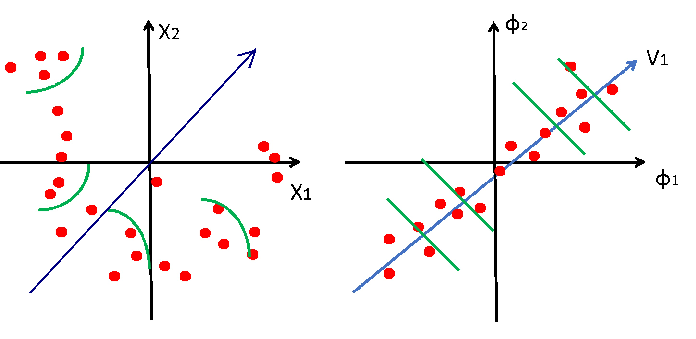}}
\caption{KPCA motivating examples: (a)  data classification with PCA (left) and KPCA (right) (b) non-linear dimension reduction of a non-linear dataset with PCA (left) and KPCA (right)}\label{fig:kpca-ex}
\end{figure}

For the sake of completeness, we include a brief matrix  derivation of  KPCA below.  More
comprehensive derivations can be found in Sch\"{o}lkopf \cite{Scholkopf:1998:NCA:295919.295960, Schölkopf1997} and Sarma \cite{Sarma2008}. Let  $N_R$ be a positive integer representing the
dimension of the random field (in this case it is equal to the number of mesh grid points),  and  $M$
be the number of observations of the random field. Given a set of discrete realizations $\{\bfy_l\}_
{l=1}^M$ of the random field where each component (or \emph{snapshot}) is $\bfy_l = [y_{1,l}, \dots, y_{N,l}]^T \in
\bbR^N, \:  l= 1,2,\dots, M$, we define a  linear or nonlinear
mapping $\mathit{\Phi}$ as:
\begin{equation}
\mathit{\Phi}:  \bbR^{N_R} \rightarrow \bbR^{N_F},\;   y_l\rightarrow \mathit{\Phi}(y_l) \in \bbR^{N_F},\; l =1,2,\dots, M,
\end{equation}
where $\bbR^{N_F}$ is the new induced feature space. Here, $N_F\gg N_R$, and the feature space $\bbR^{N_F}$ in general contains much more information (that is, higher dimension) than the original space $\bbR^{N_R}$.
For convenience,  we introduce matrix notations $\bfY :=[\bfy_1, \bfy_2, \dots, \bfy_{M}]$
and $\bfPhi :=[\mathit{\Phi}(\bfy_1), \mathit{\Phi}(\bfy_2), \dots, \mathit{\Phi}(\bfy_{M})]$.
In addition, let $\mathbf{1}_{M}:= \frac{1}{M}\mathbf{1}_{N_R\times M}$ be a matrix  with all its elements equal to $\frac{1}{M}$; and let
$\tilde{\bfY} = \bfY - \bfY \mathbf{1}_{M}$ and $\tilde{\bfPhi} := \bfPhi - \bfPhi \mathbf{1}_{M}$ be the centered matrix of $\bfY$ and  $\bfPhi$, respectively.

In  classical PCA, a discrete covariance matrix~\cite{MR2036084} is obtained as
\begin{equation}
\bfC_{o} := \frac{1}{M} \sum_{l=1}^{M} \tilde{\bfy}_l  \tilde{\bfy}_l^T = \frac{1}{M}\tilde{\bfY} \tilde{\bfY}^T.
\end{equation}
Here, the set $\{\tilde{\bfy}_l\}_{l=1}^{M}$ is a centered measurement vector given by $\tilde{\bfy}_l = \bfy_l- \bar{\bfy}$, where $\bar{\bfy} = \frac{1}{M}\sum_{l=1}^{M} \bfy_l$. Similar to the continuous version of the KL expansion with given mean and covariance kernel function, the  KL expansion of the  random fields  for the discrete case can be characterized with  following equation based on Mercer's theorem:
\begin{equation}
\bfy = D_o \Lambda_o^{1/2} \bsxi + \bfY \mathbf{1}_1,
\end{equation}
where  $D_{o}$ is a matrix of eigenvectors associated with $\bfC_{o}$;
$\Lambda_o$ is a diagonal matrix of the eigenvalues of $\bfC_{o}$;
$\bsxi  = [\xi_1, \xi_2,\cdots, \xi_{N^R}]^T\in \bbR^{N_R}$ is a column random vector with statistical properties {\color{black}$<\xi_{i}\xi_{j}>_\omega= \delta_{i,j}$ and $<\xi_i>_\omega=0$ }. {\color{black}A nonlinear choice for the $\itPhi$ such as radial basis functions leads to the nonlinear form of  PCA.  Next, we compute the centralized form of the feature vectors $\{\tilde{\itPhi}, (\bfy_l)\}_{l=1}^{M}$ where $\tilde{\Phi}(\bfy_l)= \itPhi(\bfy_l)- \bar{\itPhi}$,  $\bar{\itPhi} = \frac{1}{M}\sum_{l=1}^{M} \itPhi(\bfy_l) $.
Similar to PCA, we have the following  discrete covariance after the nonlinear mapping
\begin{equation}\label{eq:FeatureCovariance}
\bfC_{f} = \frac{1}{M} \sum_{l=1}^{M} \tilde{ \itPhi}( \bfy_l ) \tilde{ \itPhi}(\bfy_l)^T =  \frac{1}{M} \tilde{ \bfPhi} \tilde{ \bfPhi}^T.
\end{equation}

Since $N_F$ is usually much larger than $N_R$, it is infeasible in practice to perform PCA on the feature space due to the  very high dimensionality of the covariance matrix. For instance, for the polynomial kernel $(\bfx \cdot \bfy)^d$ of order $d$,  the dimension of the feature space will be~\cite{Scholkopf:1998:NCA:295919.295960}
\begin{equation}\label{eq:NF}
N_F = \frac{(N_R + d - 1)!}{ d! (N_R -1)!}.
\end{equation}
Alternatively,  the nonlinear mapping can be seen as a kernel map, thus allowing us to handle the
high dimensionality by using a technique called a ``kernel trick." A kernel trick introduces
a virtual mapping $\itPhi$, from beginning to the end, where the mapping $\itPhi$ only
acts as an intermediate functional,   resulting in smaller dimensional equivalent system compared to $\bfC$.  The
eigen-problem of the covariance matrix $\bfC_f$ in the feature space is now given as:
\begin{equation}\label{eq:kernelEigenProblem1}
\bfC_f \bfV_f = \bfV_f\Lambda_f.
\end{equation}
Here, $\bfV_f$ is the matrix of eigenvectors and  $\Lambda_f$ is a diagonal eigenvalue matrix.
The relationship between the eigenvectors $\{\bfv_l\}$ of $\bfV_f$  and the data set of $\{\tilde{\itPhi}(\bfy_l)\}$,  can be written as
\begin{equation}\label{eq:EigenEquation}
\bfC_f   \bfv_l= \frac{1}{M} \sum_{j=1}^{M} \tilde{ \itPhi}( \bfy_j ) \tilde{ \itPhi}(\bfy_i)^T, \; \:\: \bfv_l=
\frac{1}{N_R} \sum_{j=1}^{M} ( \tilde{ \itPhi}(\bfy_i)^T \bfv_l)  \tilde{ \itPhi}  ( \bfy_j )  = \gamma_l \bfv_l,
\end{equation}
which shows that the eigenvectors $\{\bfv_l\}$ are elements  in the space spanned by $\tilde{\itPhi}(\bfy_l),\: l=1,\dots, M$.

Let  $\bsalpha = [\bsalpha_1, \dots, \bsalpha_{M}]$ with $\bsalpha_l = [\alpha_{l,1}, \alpha_{l,2}, \dots, \alpha_{l,N_R}]^T$,
and  eigenmatrix $\bfV_f = \tilde{\bfPhi} \bsalpha$
where each component of the eigenvector  $\bfv_l = \sum_{j=1}^{N_R}\alpha_{l,j} \tilde{\itPhi}(y_i) = \tilde{\bfPhi} \bsalpha_l$.
Substituting this  into  Equation~\eqref{eq:EigenEquation} leads to
\begin{equation}
\bfC_f \tilde{\Phi} \bsalpha = \tilde{\Phi} \bsalpha \Lambda_f.
\end{equation}
Using the definition of $\bfC_f$ from Equation~\eqref{eq:FeatureCovariance} and
multiplying  both sides by $\tilde{\Phi}^T$, and further setting
 $K_c =    \tilde{\Phi}^T \tilde{\Phi}$,  we have
\begin{equation}
\frac{1}{M}K_c^2 \bfV= K_c  \bsalpha \Lambda_f.
\end{equation}
Assuming  $K_c$ is a nonsingular matrix, the equation above is equivalent to the following kernel eigenvalue problem
\begin{equation}\label{eq:kernelEigenProblem2}
\frac{1}{M}K_c \bfV= \bsalpha \Lambda_f,
\end{equation}
where $K_c$ is a matrix of $M\times M$.
This kernel trick  allows us to perform KPCA in the high dimensional feature space, with similar computational expense as PCA. We just need to perform an eigen-decomposition on a relatively small space $\bbR^M$, which is independent of the selection of the nonlinear mapping and  the feature space.

Solving  Equation~\eqref{eq:kernelEigenProblem2} leads to the  eigenvector matrix $\bfV$,
and the corresponding  $\bfV_f$ in Equation~\eqref{eq:kernelEigenProblem1} can be retrieved using,
\begin{equation}
\bfV_f = \tilde{\bfPhi}\bfV.
\end{equation}
Here, $\bfV_f$  has the property that
\begin{equation}
\bfV_f^T\bfV_f = \bfV^T \tilde{\Phi}^T \tilde{\Phi}  \bfV = \bfV^TK_c\bfV =M \Lambda_f.
\end{equation}
Using the same notation of $\bfV_f$, we have the orthonormal eigenvector matrix
\begin{equation}
\bfV_f = \frac{1}{\sqrt{M}}\tilde{\Phi}\bfV \Lambda_f^{-1/2}.
\end{equation}
Assuming $K = \Phi^T \Phi$,  the centered $K_c$ can be easily obtained using
 \begin{align*}
K_c &= (\Phi-\bar{\Phi})^T  (\Phi-\bar{\Phi}) = (\Phi - \Phi 1_{N_R})^T (\Phi - \Phi 1_{N_R})\\
&=\Phi^T\Phi  - \Phi^T \Phi \mathbf{1}_{N_R} - \mathbf{1}_{N_R}^T \Phi^T\Phi
 +  \mathbf{1}_{N_R} \Phi^T\Phi  \mathbf{1}_{N_R}\\
&= K - K \mathbf{1} - \mathbf{1} K + \mathbf{1}K\mathbf{1}
\end{align*}
Thus, we have  the KL expansion in the feature space as
\begin{align} \label{eq:kernelKL}
\bfY_f =    \bfV \Lambda^{1/2} \bsxi + \bar{\Phi}=  \frac{1}{\sqrt{M}}\tilde{\Phi}\bfV \Lambda_F^{-1/2}\Lambda^{1/2}\bsxi + \bar{\Phi}
= \frac{1}{\sqrt{M}}\tilde{\Phi} \bfV \bsxi + \bar{\Phi},
 \end{align}
where $\bsxi = [\xi_1,   \dots, \xi_{N_R}]^T$ is a random vector with properties  $\bbe[\xi_i] = 0, \bbe[\xi_i \xi_j] = \delta_{i,j}$.
The polynomial kernel and Gaussian kernel defined below are frequently used in
practice, which are given by
\begin{eqnarray}
 &k(\bfx, \bfy) = c+ (\bfx\cdot \bfy)^d , \;d\geq 1,\\
& k(\bfx, \bfy) = \exp(-\frac{\norm{\bfx-\bfy}^2}{\sigma}),\; \sigma > 0,
\end{eqnarray}
respectively. Kernel functions directly calculate  the dot product in the space of $\bbR^F$ using
elements in the input space $\bbR^{N_R}$. Since there is no actual mapping of $\itPhi(y)$, kernels play
the role of the intermediate functional.

Although stochastic inversion is performed in the feature space, our interest is to obtain the {\color{black} snapshots}  from the posterior  in the original space $\bbR^{N_R}$. In order to achieve this, a pre-imaging problem is solved to project snapshots from the feature space back to the original space. In general, due to the non-linearity of the mapping $\itPhi$, neither existence nor uniqueness of the pre-image is guaranteed.
One method to perform pre-imaging involves solving the following optimization problem
~\cite{Scholkopf:1998:NCA:295919.295960},
\begin{equation}
\min_{\bfy} \rho(\bfy) = \norm{\Phi(\bfy) - \bm Y}^2,
\end{equation}
where the  $\bfy \in \bbR^{N_F}$ and $\bm Y \in \bbR^{N_{R}}$ are points in the feature space
and original space, respectively, and $\|\cdot \|$ is the Euclidean norm.
The above minimization problem can be reduced to the following  iterative fixed point problem~ \cite{Sarma2008, Scholkopf:1998:NCA:295919.295960}
\begin{equation}\label{preimageiteration}
\bfy^{k+1} = \frac{\sum_{l=1}^{N_R} \beta_i  \sum_{j=1}^d    j(\bfy_i \cdot y^k)^{j-1} \bfy_i}{  \sum_{l=1}^{N_R} \beta_i  \sum_{j=1}^d    j(\bfy_i \cdot y^k)^{j-1}  }.
\end{equation}
Note here that non-iterative pre-imaging techniques based on reproducing kernel Hilbert space (RKHS) also developed by several researchers~\cite{kwok2003pre,bengio2004out,arias2007connecting,honeine2009solving} and a comprehensive comparison these methods can be found in~\cite{honeine2011preimage}.

The resulting KPCA method allows us to find a low-dimensional, relevant feature space and obtain a pre-image. The next section introduces a procedure to efficiently sample the KPCA-feature random variables.
\subsection{Mapping  non-Gaussian feature random variables to Gaussian random variables}\label{sec:map}
KPCA feature random variables are uncorrelated but dependent non-Gaussian random variables. This section introduces a ICDF-transformation-based PCE construction to sample from the feature random variables.

Let $\bsxi^d$ be the discrete observations of  $\bsxi$ obtained from the measurements
of the snapshots $\{\bfy_l\}_{l=1}^{M}$. Letting $\bfY_f = \bfPhi$
and multiplying both sides of Equation~\eqref{eq:kernelKL} by $\bfPhi^T$, we obtain
\begin{equation}
\tilde{\bfPhi} = \bfPhi - \bfPhi \mathbf{1}_{M}  = \ \frac{1}{\sqrt{M}}\tilde{\Phi} \bfV \bsxi^d
\Rightarrow K_c = \frac{1}{\sqrt{M}} K_c \bfV\bsxi^d.
\end{equation}
Assuming $K_c$ is nonsingular, we have
\begin{equation}\label{discrete_xi}
\bfV\bsxi^d = \sqrt{M}\mathbf{1}_M
\end{equation}
which can be solved using a least-squares method or singular value decomposition (SVD).

Random variables $\bsxi^d$ computed from Equation~(\ref{discrete_xi}) act as a prior
distribution for the Bayesian inversion framework. In general, $\bsxi^d$ are
non-Gaussian, uncorrelated and dependent random variables, which may complicate the Bayesian inversion procedures
(e.g. by requiring more frequent sampling from their distributions).

Determination of a unique map from the dependent $\bsxi^d$ to a standard independent random variable space
$\bseta$ is an active research area. One way to achieve a non-unique
mapping is using iso-probabilistic mappings such as the generalized Nataf transformation~\cite
{lebrun2009generalization}  and Rosenblatt transformation~\cite{rosenblatt1952remarks}. However, these transformations require information such as conditional  distributions,
which are hard to construct from limited observations. Therefore, we assume $\{\xi{_l^d}\}_{l=1}^{M}$ are
independent similar to \cite{ghanem2006construction,stefanou2009identification}, and to
facilitate the sampling we construct  a  polynomial chaos expansion (PCE) for each {\color{black}$\xi_l^d$}.

PCE, originally introduced by Wiener~\cite{wiener1938homogeneous,ghanem2003stochastic}, represents
any  random variable with finite variance as a summation of a series of polynomials over the centered
normalized Gaussian variables. We can represent each component of  $\{\xi{_l^d}\}_
{l=1}^{M}$ obtained from Equation~\eqref{discrete_xi} using  PCE as
\begin{equation}\label{eq:onedimensionalPCE}
 \xi{_l^d} =  \sum_{n=0}^{\infty} c_{n,l} \Psi_{n}(\eta_l(\omega)),  l = 1,2, ...,
\end{equation}
where $\eta_l$ are i.i.d. standard Gaussian random variables,
$\Psi_{n}(\eta_l(\omega))$ are Hermite polynomials, and $c_{n,l}$ are real valued deterministic coefficients.
The associated orthogonal system $\{ \Psi_{n}(\eta) \}_{n\in \mathbb{N}}$ forms the
homogeneous polynomial chaos basis. The
coefficients in the equation above can be computed using Bayesian inference~\cite
{arnst2010identification} or using a non-intrusive projection method~\cite
{eldred2009comparison}. We use a  projection method~\cite{NME:NME2546} to find a continuous parameterized
representation similar to Equation~\eqref{eq:onedimensionalPCE} based on the
discrete $\bsxi^d$. Let $\{\eta_l\}$ be a standard Gaussian random variable, then by
matching the cumulative density function (cdf) of $\xi{_l^d}$ and $\eta_l$,   each
component of $\xi_l$ can be expressed in terms of random variables $\eta_l$ by  following
 non-linear mapping:
\begin{equation}\label{eq:xi_eta}
 \xi{_l^d} =  F_{\xi{_l^d}}^{-1}\circ F_{\eta_l} (\eta_l),
\end{equation}
where $F_{\xi{_l^d}}$ and  $F_{\eta_l}$ denote the cdfs of $\xi{_l^d}$ and $\eta_l$
respectively.
The coefficients of the PCE are then computed using the projection of
$F_{\xi{_i^d}}^{-1} \circ F_{\eta_l}$ on the orthonormal chaos basis system,
\begin{equation}
  c_{n,l} = <\xi{_l^d}, \Psi_n> = \int_{\Omega} F_{\xi{_l^d}}^{-1}\circ F_{\eta_l} \Psi_n d\bbP_\eta(\omega),
\end{equation}
However, the {\color{black}cdf} $F_{\xi{_l^d}}$ is not known and needs to be estimated using the empirical cdf \cite{10.2307/2242455} based on the discrete observations of $\bsxi^d$.
 The empirical {\color{black}cdf} ($\tilde{F}_{\xi{_l^d}}$) of $\xi{_l^d}$ can be estimated from sampling using,
\begin{equation}
  \tilde{F}_{\xi{_i^d}}(x) = \frac{1}{M} \sum_{k=1}^{M} I (\xi{_l^d}^{(k)} \leq x),
\end{equation}
where $I(A)$ is the indicator function of event $A$.  We then introduce the
 following approximation
\begin{equation}
  F_{\xi{_i^d}}^{-1} \sim \tilde{F}_{\xi{_i^d}}^{-1}, \: \text{where} \: \tilde{F}_{\xi{_i^d}}^{-1} :[0,1]\rightarrow \bbR
\end{equation}
which is uniquely defined as
\begin{equation}
   \tilde{F}_{\xi{_i^d}}^{-1}(y) = \min\{x \in \{ \xi_{l^d}^{(k)}\}_{k=1}^M; \tilde{F}_{\xi{_i^d}}(x) \geq y\}.
\end{equation}
Then the coefficients of the  polynomial chaos expansion can be computed  using a numerical
integration. Instead of using the indicator functions,  we use kernel density
estimation~\cite{jones1990performance} to construct the empirical {\color{black}cdf},
\begin{equation}
\tilde{f(\xi)} = \frac{1}{M} \sum_{l=1}^{M} {\color{black} K_h} (\xi - \xi_l),
\end{equation}
where $K_h(\cdot)$ is the kernel function.
\begin{equation}
c_{n,l} = <\xi_l, \Psi_n> = \int_{\Omega} F_{\xi_l^d}^{-1}\circ F_{\eta_l^d} \Psi_n d\bbP_\eta(\omega),
 = \int_{\Omega} F_{\xi_l^d}^{-1}\circ F_{\eta_l^d} \Psi_n \frac{e^{-\eta^2/2}}{\sqrt{2\pi}} dx
\end{equation}
The coefficients $c_{n,l}$  can be efficiently calculated using the Gauss-Hermite quadrature rules.

The above procedure allows us to sample from the feature random variables within the Bayesian inference framework.

\subsection{Bayesian inference}
Bayesian inference provides a systematic framework for integrating prior knowledge and measurement uncertainties and computes a probabilistic solution to the inverse problem. It treats the parameters $\mu(\bm x), \: \lambda(\bm x)$ of  the forward
model~\eqref{eq:model} as a random process. Instead of performing Bayesian
inference with respect to these parameters directly, we perform the inference in the extracted feature space of $\bseta$.
We denote the stochastic elasticity forward model \eqref{eq:model} as {\color{black} $\bfu = f
(\bseta)$}, which describes the relationship between the observed output state $\bfu_{obs}$ and the
uncertain model parameters $\bseta$. As such, the posterior distribution from the Bayesian inference can be expressed as
\begin{equation}\label{eq:BaysianInference}
\pi_{posterior}(\bseta) : = \pi (  \bseta | \bfu_{obs} ) \propto   \pi_{prior}(\bseta)  \pi_{likelihood}(\bfu_{obs}| \bseta) {\color{black}.}
\end{equation}
This approach allows us to fuse simulations and measurements into the inversion framework. Unlike deterministic inversion,
the expression \eqref{eq:BaysianInference} provides a probabilistic characterization of the solution~\cite{martin2012stochastic} for the {\color{black} stochastic} inverse problem. In this context, the likelihood function $\pi_{likehood} (\bfu_{obs}| \bseta)$ is a conditional probability of the model outputs with given model parameters $\bseta$. Also, the prior probability density function {\color{black}(pdf)}  $\pi_{prior}(\bseta)$ allows us to inject prior knowledge into the model. In our case,
the prior density function $\pi_{prior}$ is a  multivariate Gaussian of the form:
\begin{equation}\label{eq:priorpdf}
\pi_{prior} (\bseta) \propto \exp (     -\frac{1}{2}  { \|\bseta - \bar{\bseta} \|}_{\Gamma_{prior}^{-1}}^2).
\end{equation}
The simplification above is possible due to the independence of the $\bseta$ vector. Specifically, the covariance matrix $\Gamma_{prior}$  is an identity matrix and
$\bar{\bseta}$ is a zero vector.
The representation of likelihood function is core to the characterization of the posterior density function $\pi_{posterior}$.
In the limiting case where the measurement and the model are exactly unbiased, the Bayesian model can easily be reduced to
\begin{equation} \label{eq:bayesiansolution}
\pi_{posterior}(\bseta) : = \pi (  \bseta | \bfu_{obs} ) \propto   \pi_{prior}(\bseta).
\end{equation}
To further simplify the discussion, here we assume that the error between the measurement and the model is unbiased and additive,
and the noise follows a Gaussian distribution. This leads to following expression for the likelihood function
\begin{equation}\label{eq:lik}
\pi_{likelihood}(\bfu_{obs}| \bseta) \propto \exp (     -\frac{1}{2}  { \|f(\bseta)-  \bfu_{obs} \|}_{\Gamma_{noise}^{-1}}^2).
\end{equation}
We note that our procedure is still valid for other choices of likelihood functions. Our particular choice for likelihood is due to limited information on  measurement and modeling errors. The choice of the likelihood function of the form Equation~(\ref{eq:lik}) leads to following log-likelihood function,
\begin{equation}
-\log(  \pi(\bfu_{obs}| \bm\eta)   ) = \frac{1}{2}  { \|f(\bseta)-  \bfu_{obs} \|}_{\Gamma_{noise}^{-1}}^2,
\end{equation}
and  the corresponding posterior density can be derived as
\begin{equation}
\pi_{posterior}(\bseta) \propto \exp ( J(\bseta)),
\end{equation}
where  $J(\bseta)$ is given by
{\color{black}
\begin{equation}
J(\bseta)  : =  \frac{1}{2}  { \|f(\bseta)-  \bfu_{obs} \|}_{\Gamma_{noise}^{-1}}^2 +\frac{1}{2}  { \|\bseta - \bar{\bseta} \|}_{\Gamma_{prior}^{-1}}^2.
\end{equation}
}

Due to the non-linear relation between the parameters $\bseta$ and the measurements, direct sampling from the posterior is not possible even with the chosen likelihood function~\cite{martin2012stochastic}. MCMC methods provide a systematic way to sample from the corresponding posteriors.

\subsection{Gradient-based adjoint MCMC}

The nonlinear mapping between the observables and parameters leads to non-Gaussian posteriors even with additive noise and a Gaussian prior assumption.  MCMC methods are relevant techniques for sampling non-standard posteriors. They require many simulations of the forward models, however, leading to computational intractability when the forward models are expensive to evaluate. Here, we employ LMCMC to reduce the computational complexity, using gradient information computed in the feature space based on the adjoint PDE and automatic differentiation in the feature space. Theoretically, LMCMC has a computational complexity of $O(n^{1/3})$, while Metropolis Hastings MCMC (MHMCMC) based on random walk has the complexity of $O(n)$ where $n$ is the dimension of the inference parameters. LMCMC considers the following overdamped Langevin-Ito diffusion process,
\begin{equation}
dX = \nabla \log \pi_{posterior} (X)dt  + \sqrt{2} dW.
\end{equation}
The probability distribution $\rho(t)$ of $X(t)$ approaches a stationary distribution, which is invariant
under diffusion, and  $\rho(t)$  approaches the true posterior ($\rho_{\infty} = \pi_{poster}$) asymptotically.
Approximate sample paths of the Langevin diffusion can be generated by many discrete-time methods. Using a fixed time step $\tau >0 $, the above equation can be written as,
\begin{equation}
X_{k+1} = X_k + \tau \nabla \log \pi (X_k) + \sqrt{2 \tau} \xi_k
\end{equation}
where each $\xi_k$ is an independent draw from a multivariate normal distribution on $\bbR^{N_F}$ with mean $0$ and identity covariance matrix.

This proposal is accepted or rejected similar to the Metropolis-Hasting algorithm using $\alpha$,
\begin{equation}
\alpha = \min \{1,        \frac{\pi (X_{k+1})  q(X_k | X_{k+1}) }{  \pi( X_k)   q (X_{k+1}   | X_k )} \}
\end{equation}
where
\begin{equation}\label{eq:qx}
q(x'|x) \propto \exp (    -\frac{1}{4\tau}   \| x' -x - \tau \nabla  \log \pi(x)   \|_2^2 )
\end{equation}

\subsection{Adjoint Information of the posterior density function }
In this section, we introduce a technique to compute the gradient information of the negative logarithm of the posterior function with respect to the random parameters $\bseta$,
\begin{eqnarray}
J(\bseta)  &: =  \frac{1}{2}  { \|f(\bseta)-  \bfu_{obs}\|}_{\Gamma_{noise}^{-1}}^2 +\frac{1}{2}  { \|\bseta - \bar{\bseta} \|}_{\Gamma_{prior}^{-1}}^2 \\
           &= J_1(\bseta) + J_2(\bseta), \label{eq:v1}
\end{eqnarray}
where $J_1(\bseta)= \frac{1}{2}  { \|f(\bseta)-  \bfu_{obs}\|}_{\Gamma_{noise}^{-1}}^2$
and $ J_2(\bseta)=\frac{1}{2}  { \|\bseta - \bar{\bseta} \|}_{\Gamma_{prior}^{-1}}^2$.
It is nontrivial to obtain the functional derivative of $J(\bseta)$. Here we use the adjoint model and automatic differentiation to compute the gradients. Using
the mathematical derivations in the preceding sections, the relationship between the variables $\bseta,\bsxi,\bfy, \mu , \lambda, \bfu$ can be summarized as,
\begin{equation}
\bseta \xrightarrow{ \text{PCE}} \bsxi    \xrightarrow{ \text{Pre-image}}  \bfy \xrightarrow {\text{exp}} \mu , \lambda \xrightarrow{\text{forward\;model}} \bfu.
\end{equation}

The objective functional $J$ can be expressed in terms of  $\bseta$ by
\begin{eqnarray}
\bseta \rightarrow  \frac{1}{2} (  f(\bseta) - \bfu_{obs}, \Gamma_{noise}^{-1}(f(\bseta) - \bfu_{obs}) ) + \frac{1}{2} (\bseta - \bar{\bseta},\Gamma_{prior}^{-1}( \bseta - \bar{\bseta})  )
\end{eqnarray}

The second part of $J(\bseta)$ is a quadratic form in the parameters $\bseta$. The expression for the gradient of $J_2(\bseta)$ can directly be obtained as
\begin{equation}
\nabla_{\bseta} J_2(\bseta) =  \Gamma_{prior}^{-1}( \bseta - \bar{\bseta} )
\end{equation}

To derive the  gradient of  $J_1$, we follow the procedure similar to Giering {\it et al.} \cite{giering1998recipes}.
Consider the Taylor expansion $J_1$ with respect to the control variables at a given point $\bseta_0$
\begin{equation}
J_1(\bseta) = J_1(\bseta_0) + (\nabla_{\bseta} J_1(\bseta_0), \bseta - \bseta_0) + O(\abs{\bseta-\bseta_0}),
\end{equation}
or in shorthand,

\begin{equation}
\delta J_1  = (\nabla_{\bseta} J_1(\bseta_0), \delta \bseta).
\end{equation}
We use the shorthand notation whenever linear approximations are involved. Suppose $J_1$ is sufficiently regular, then for each parameter vector $\bseta_0$, and using symmetry property of the inner product and applying the product rule of differentiation  yields
 \begin{equation}
 \delta J_1 = (\Gamma_{noise}^{-1}(f(\bseta) - \bfu_{obs}), \nabla_{\bseta} f(\bseta_0) \delta \bseta).
 \end{equation}
Using the definition of the adjoint operator we obtain
\begin{equation}
\delta J_1 = ( (\nabla_{\bseta} f(\bseta_0))^T \Gamma_{noise}^{-1}(f(\bseta) - \bfu_{obs}), \delta \bseta).
\end{equation}
Therefore, according to the definition of gradient, the gradient of the $J_1$ with respect to $\bseta$ is
\begin{equation}
\nabla_{\bseta} J_1 ( \bseta_0 ) = (\nabla_{\bseta} f(\bseta_0))^T \Gamma_{noise}^{-1}(f(\bseta) - \bfu_{obs}),
\end{equation}
Since  the function $f : =   f_1 \circ f_2 \circ f_3 \circ f_4$, applying the chain rule yields
\begin{align}
f' : &=   f_1' \circ f_2' \circ f_3' \circ f_4' \\
       &{\color{black} = \nabla_{\lambda,\mu} \bfu \nabla_{\bfy}\lambda \nabla_{\bsxi} \bfy
        \nabla_{\bseta}\bsxi.}
\end{align}

The gradient information can be rewritten as
\begin{equation}
\nabla_{\bseta} J_1 ( \bseta_0 ) =
(\nabla_{\bseta}\bsxi)^T
(\nabla_{\bsxi} \bfy)^T
(\nabla_{\bfy}\lambda)^T
( \nabla_{\lambda,\mu} \bfu )^T \Gamma_{noise}^{-1}(f(\bseta) - \bfu_{obs}),
\end{equation}
The linear operator $ \nabla_{\lambda,\mu} \bfu$ represents the tangent linear model of the forward problem and its adjoint
operator is $(\nabla_{\lambda,\mu} \bfu)^T$. Both operators depend on the point $\bseta_0$ at which the model is linearized.
The linear operator $(\nabla_{\bseta}\bsxi)^T$ represents the adjoint model of the PCE, and $(\nabla_{\bsxi} \bfy)^T$ represents the adjoint
model of the pre-image iteration mapping.

The  adjoint model {\color{black}$(\nabla_{\lambda,\mu} \bfu)^T$}  can easily be obtained with the procedure detailed in \S\ref{elasticitymodel}.  
The PCE mapping in Equation~\eqref{eq:onedimensionalPCE} and the pre-image mapping methods are continuous smooth mappings.
The adjoint models for these mappings are obtained with automatic differentiation~\cite{TapenadeRef13}.

\subsection{Algorithms}
In this section, we summarize the above derivations into two simple algorithms to facilitate the implementation of the proposed methodology.

\begin{algorithm}[H]
	\caption{Computation of posterior density function and gradients}
	\begin{algorithmic}[]
	\State {Read the snapshots $\{\bm y_l\}_{l=1}^{M}$ of the parameters  $\mu, \lambda$}
		\State {Compute KPCA reduced model using  Equation~\eqref{eq:kernelKL}}
		\State Parameterize the random variables $\bsxi$  with PCE using  Equation~\eqref{eq:onedimensionalPCE}
		\State Compute prior density function $\pi_{prior}$  as defined by  Equation~\eqref{eq:priorpdf}
		\State Compute likelihood function $\pi_{likelihood}$ as defined by  Equation~\eqref{eq:lik}
		\State { \color{black}Compute the posterior density function using  Equation~\eqref{eq:BaysianInference}   } 
		\State Compute the  gradient of the cost functional with respect to parameters $\lambda$ and $\mu$ using adjoint model
		\State Compute the  gradient of the cost functional in the feature space using automatic differentiation
	\end{algorithmic}
\end{algorithm}

\begin{algorithm}[H]
\caption{Posterior sampling using Langevin MCMC framework}
\begin{algorithmic}[]
\State Choose initial parameters $\bseta_0$
\State Compute $\pi_{posterior} (\bseta_0)$ using Algorithm 1
\For {$l$=1 to N}
\State Draw sample $y$ from the proposal density function
\State Compute $\pi_{posterior}(y)$ using algorithm 1
\State Compute $\alpha(\bseta_l, y) = \min \{ 1, \frac{\pi_{posterior}(y) q(y | \bseta_l ) }{ \pi_{posterior}(\bseta_l)   q(\bseta_l | y)}\}$, where $q(y | \bseta_l )$ and
\State $q(\bseta_l | y)$ are computed using  Equation~\ref{eq:qx}
\State Draw  $u\sim U([0,1])$
\If   {$ \;u < \alpha(\bseta_l, y)$}
\State Accept : Set $\bseta_{l+1} = y$
 \Else
\State Reject : Set $\bseta_{l+1} = \bseta_k$
\EndIf
\EndFor
\end{algorithmic}
\end{algorithm}
\FloatBarrier
\section{Numerical Simulations}\label{sec:ns}
\begin{figure}[h]
\centering
\subfloat[][]{\includegraphics[width=0.49\linewidth]{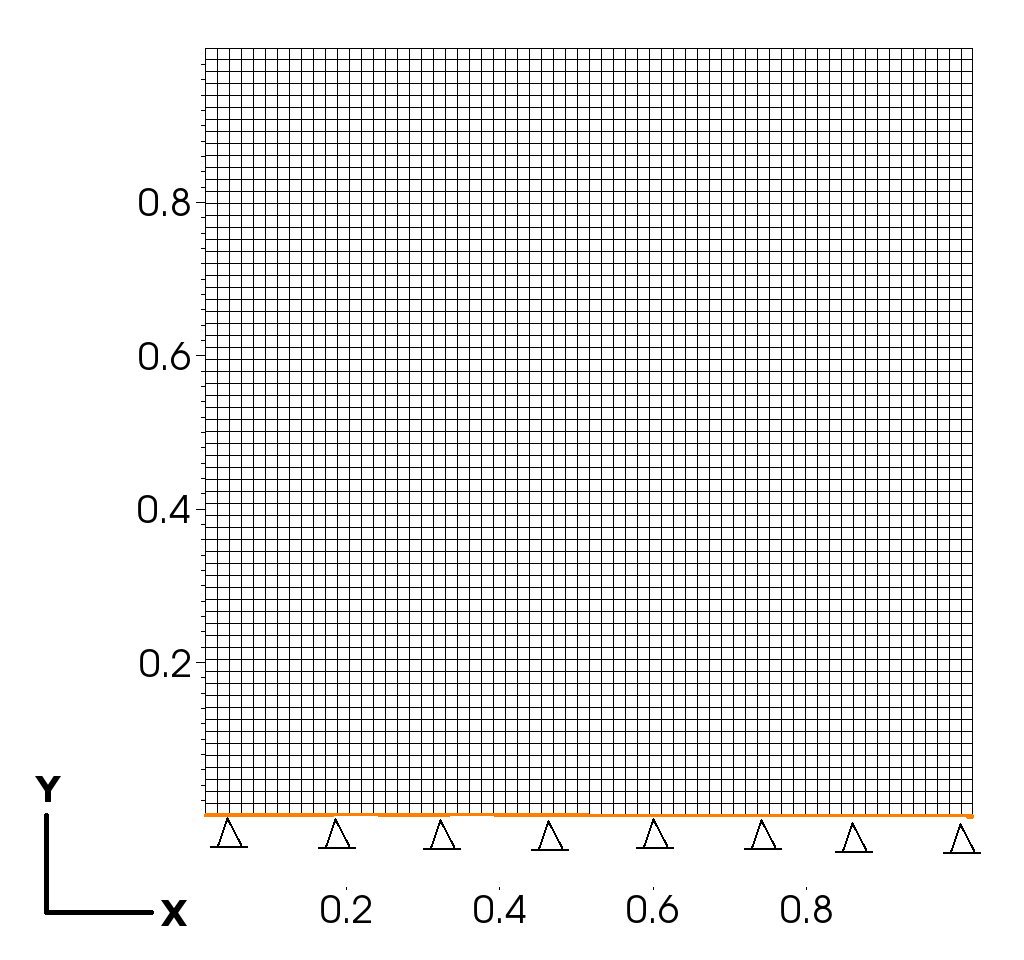}}
\subfloat[][]{\includegraphics[width=0.49\linewidth]{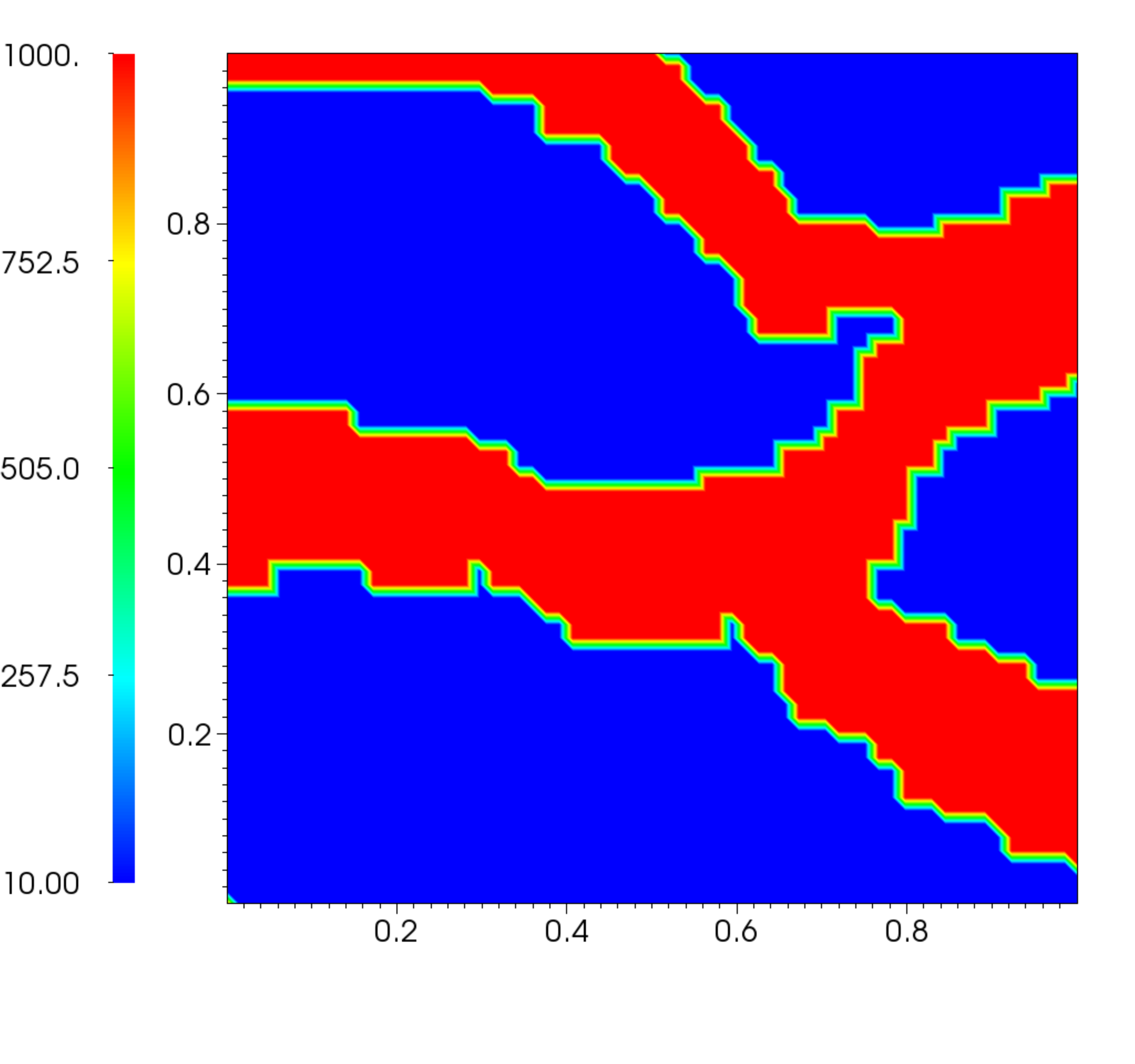}}\\
\subfloat[][]{\includegraphics[width=0.49\linewidth]{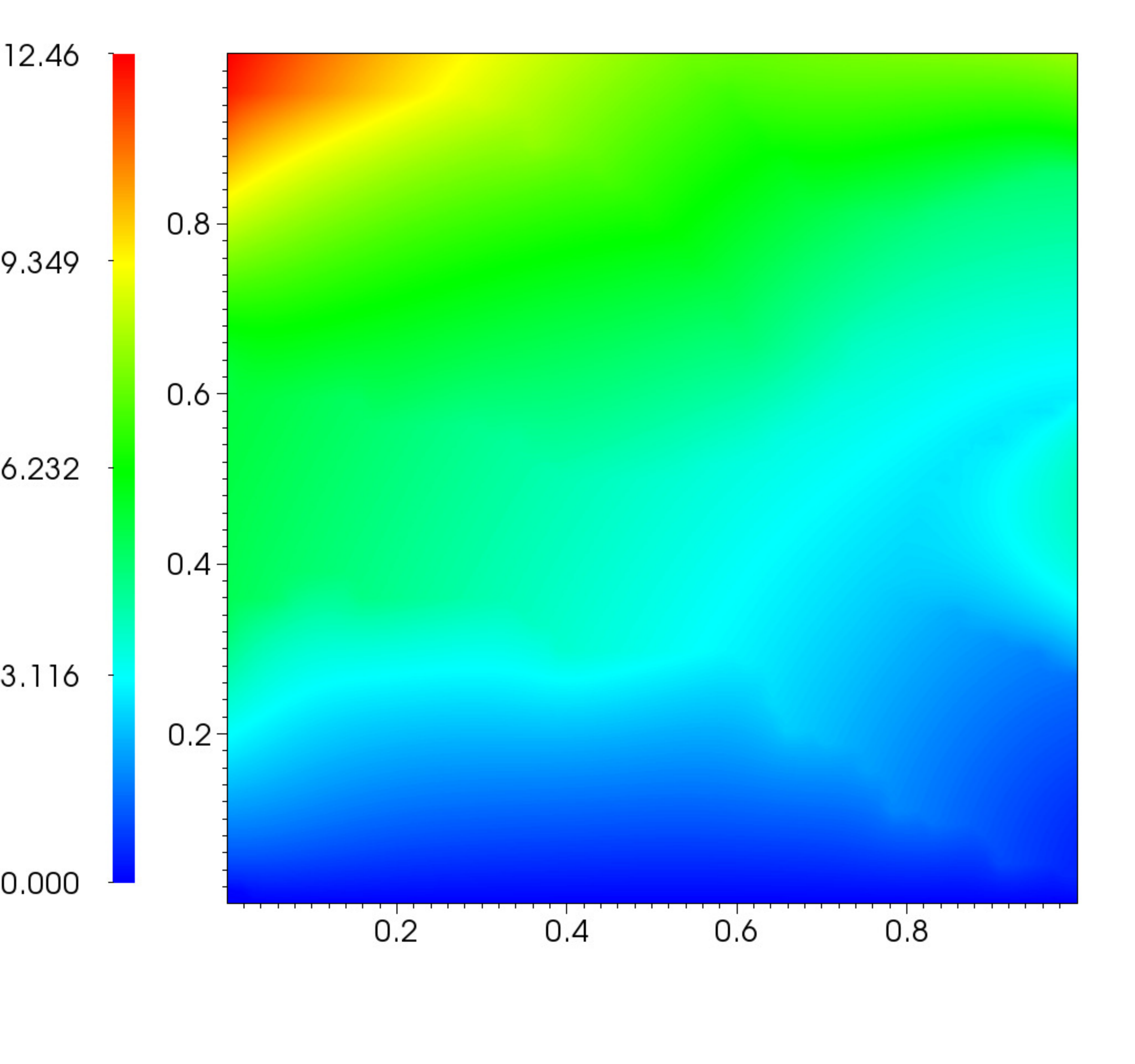}}
\caption{a) Physical setup of the numerical example used for the demonstration b) a realization $ \lambda_1$ of the elasticity parameters c) corresponding displacement magnitude due to self weight. }\label{fig:set1}
\end{figure}
In this section, we demonstrate the computational efficiency of the proposed method for the stochastic inversion of a 2D-linear elasticity model through a numerical example. The objective is to recover elasticity parameters of a geologically-complex rock characterized by sinuous channels of one material embedded in another. Figure~\ref{fig:set1} (a) shows the mesh and boundary conditions of the numerical example. The bottom boundary is supported by a pinned connection to curtail vertical and horizontal motion and other boundaries are free to expand. The square shaped domain is allowed to deform under self-weight due to gravity. Measurements of the displacements are assumed to be available at the top, left and right boundaries. For the sake of simplicity, we assume the Poisson ratio of rock is fixed at a typical value of $\nu=0.25$.  This implies $\lambda = \mu$, and therefore we need only invert for one elastic parameter field instead of two. Figure~\ref{fig:set1} (b) depicts a discrete realization $\lambda_1$. Blue and red color domains here correspond to two distinct rock types with considerable differences in their elastic properties. Homogeneous elasticity models tend to over simplify the system and can lead to sub-optimal solutions. Figure~\ref{fig:set1} (c) shows a contour plot of the displacement magnitude with elasticity parameters $\lambda_1$.
\begin{figure}[h]
\centering
\subfloat[][]{\includegraphics[width=0.48\linewidth]{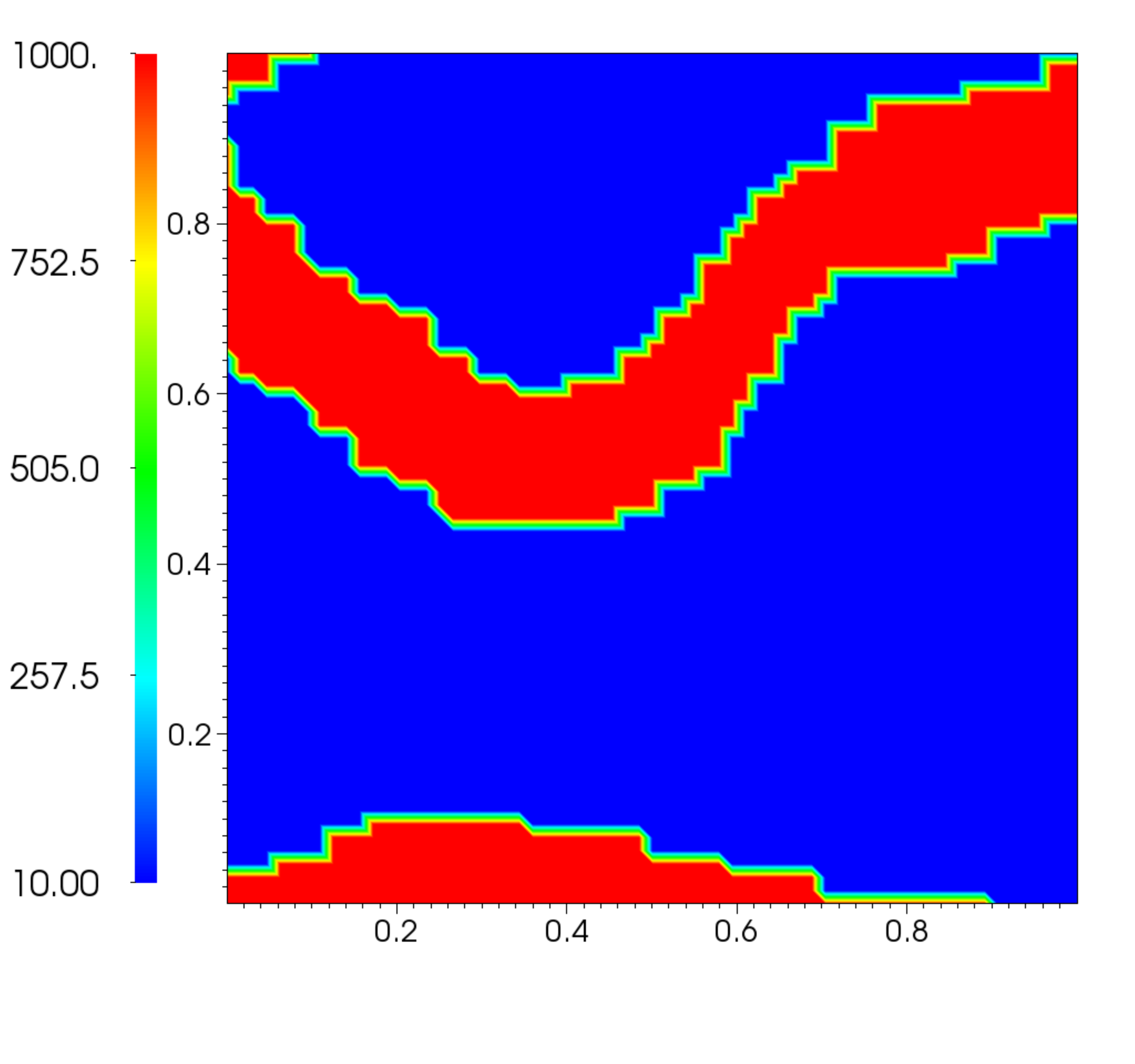}}
\subfloat[][]{\includegraphics[width=0.48\linewidth]{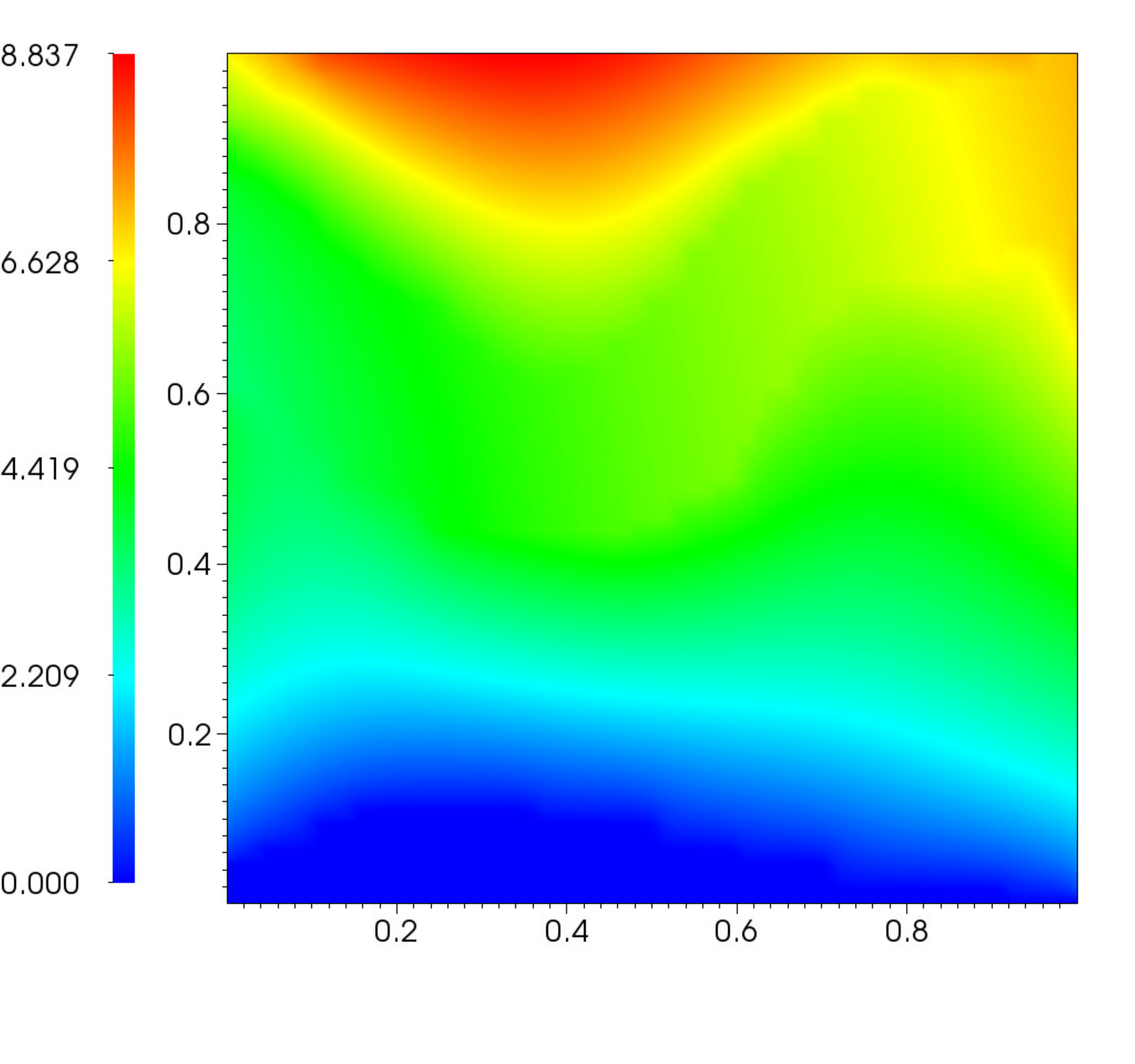}}\\
\subfloat[][]{\includegraphics[width=0.48\linewidth]{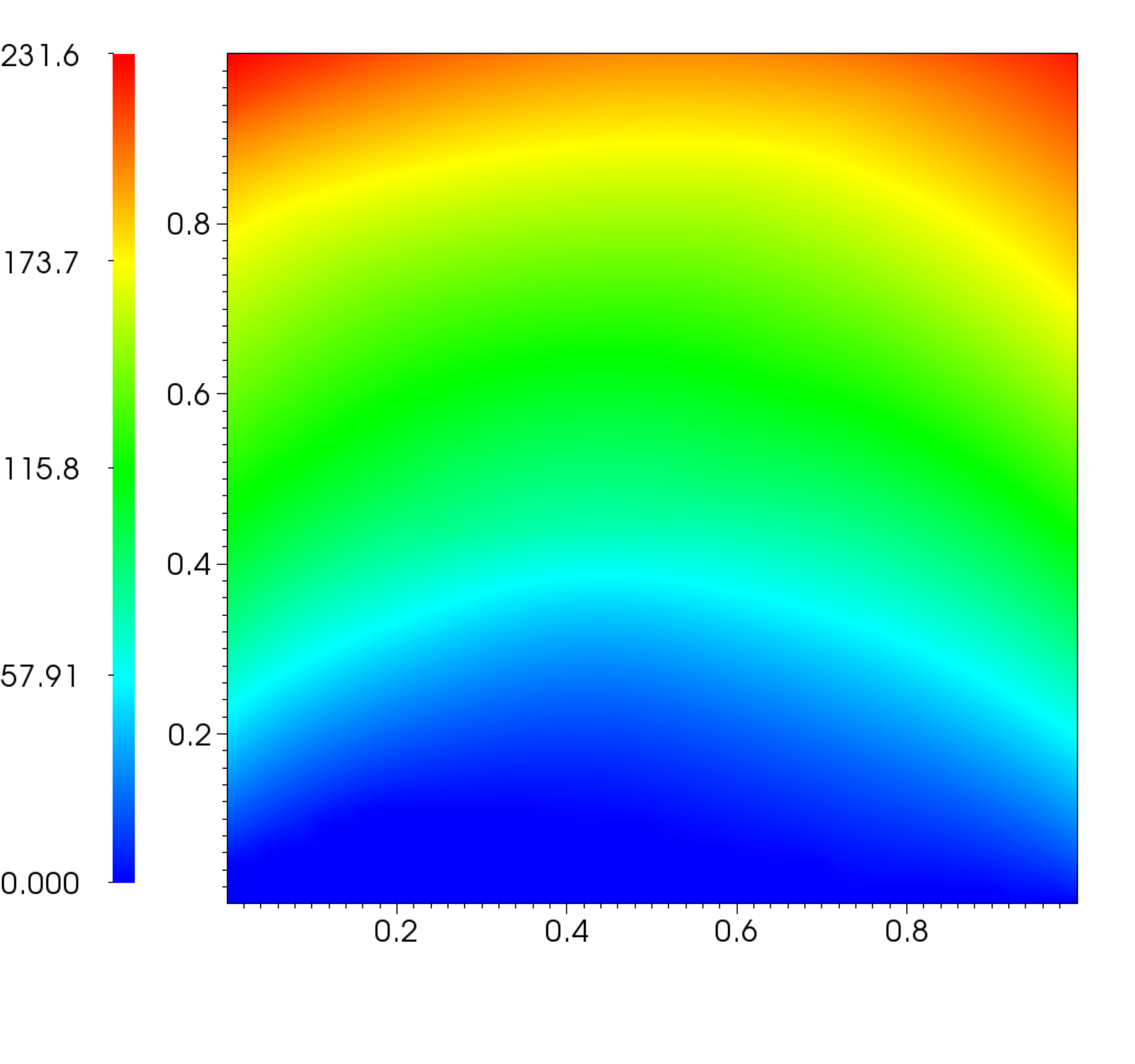}}
\subfloat[][]{\includegraphics[width=0.48\linewidth]{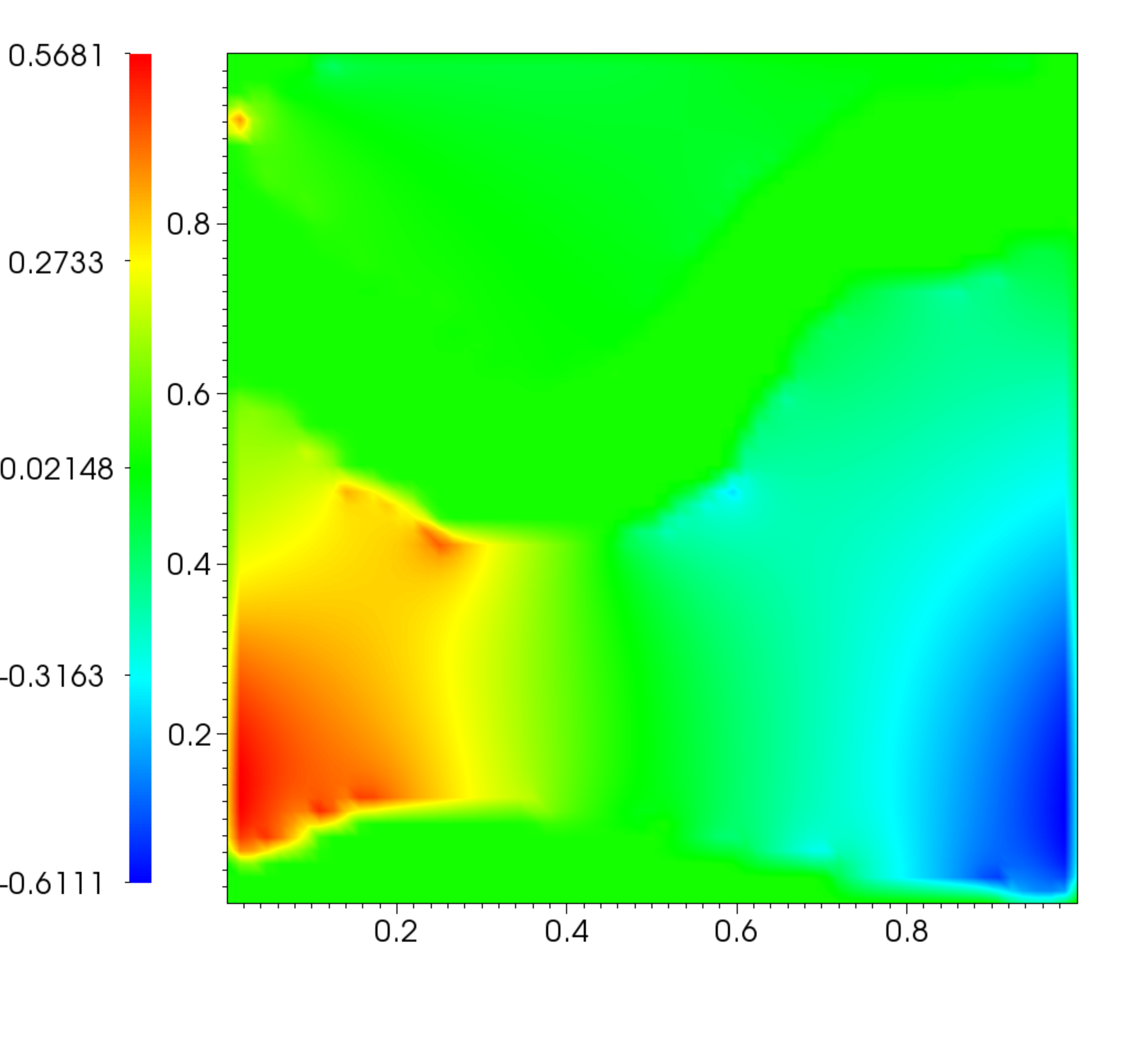}}
\caption{(a) A realization of the elasticity parameter $\lambda_2$ (b) forward displacement magnitude due to self weight (c) adjoint displacement magnitude (d)  gradient of the cost function with respect to $\lambda_2$  based on the measurements obtained with elasticity parameters $\lambda_1$ }\label{fig:set2}
\end{figure}
A forward and an adjoint simulations are performed in any LMCMC sampling step to compute the gradient of the cost functional with respect to the model parameters. Figure~\ref{fig:set2} demonstrates an example of the gradient computation, here, Fig.~\ref{fig:set2} (a) shows a realization of the elasticity parameter $\lambda_2$, used to evaluate the adjoint solution based on the measurements obtained with parameters $\lambda_1$. Figure~\ref{fig:set2} (b) depicts forward displacement magnitude of the  model with parameters $\lambda_2$ due to self weight and  Fig~\ref{fig:set2} (c) shows the corresponding adjoint displacement magnitude contour computed with the adjoint PDE.  Figure~\ref{fig:set2} (d) shows the gradient of the cost function with respect to $\lambda_2$ evaluated with self-adjoint PDE formulation. 

\subsection{Snapshot generation}
\begin{figure}[!htp]
\centering
\subfloat[][]{\includegraphics[width=0.49\textwidth]{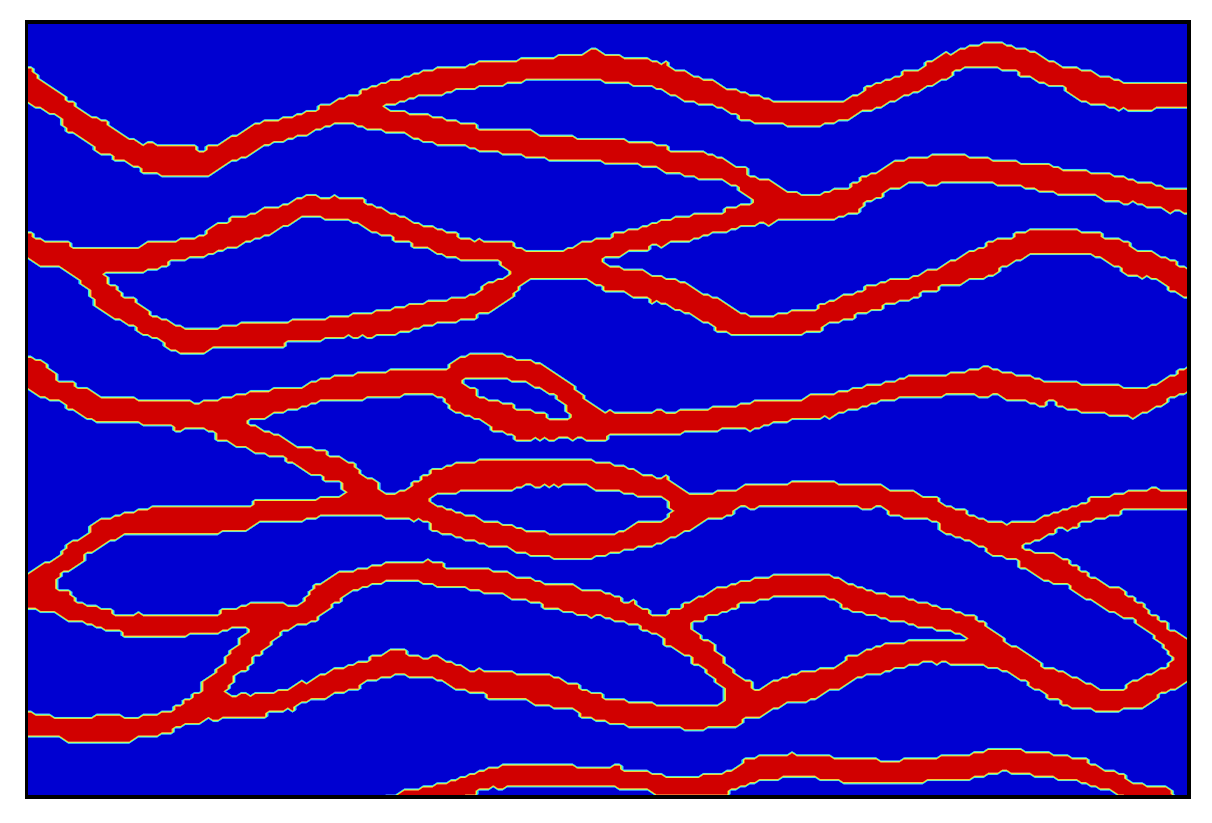}}
\subfloat[][]{\includegraphics[width=0.47\textwidth,,trim={1cm 1cm 1cm 1.1cm},clip]{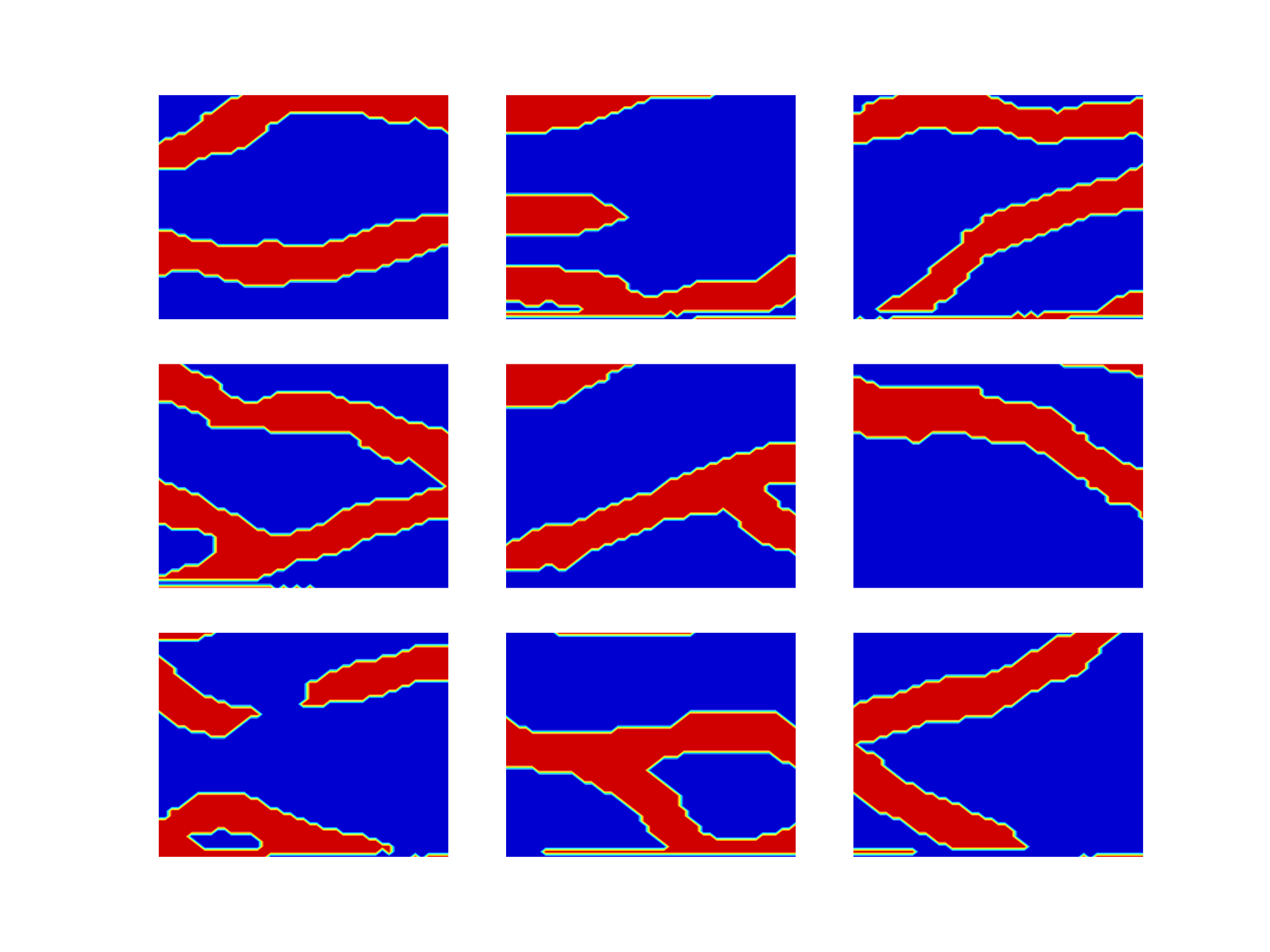}}
\caption{ (a) Training image and (b) a few snapshots generated  with the SNESIM  algorithm}\label{fig:trps}
\end{figure}
For natural materials like rock, elasticity parameters often exhibit multi-scale spatial fluctuations due to inherent heterogeneity~\cite{thimmisetty2015multiscale}.  In our numerical experiments, we rely on the single normal equation simulation (SNESIM) algorithm \cite{strebelle2002conditional} based on a training image, as shown in Fig.~\ref{fig:trps} (a), similar to Ma et al. and Sarma et al.~\cite{Ma:2011:KPC:2016171.2016423, Sarma2008}.  We generate 1000 realizations of a ``channelized" rock.  Figure~\ref{fig:trps} (right) depicts a few snapshots generated using the SNESIM algorithm. Here, $\lambda$ for the channel material (red) and host material (blue) are assumed to be 10 and 1000 MPa, respectively.  In order to guarantee positive values for the elasticity parameters, the inversion procedure is carried on  $\ln(\lambda)$.
\subsection{ Efficiency of the kernel PCA and the pre-image}
\begin{figure}[h]
\centering
\subfloat[][]{\includegraphics[width=0.46\linewidth,trim={7cm 6cm 0 0},clip]{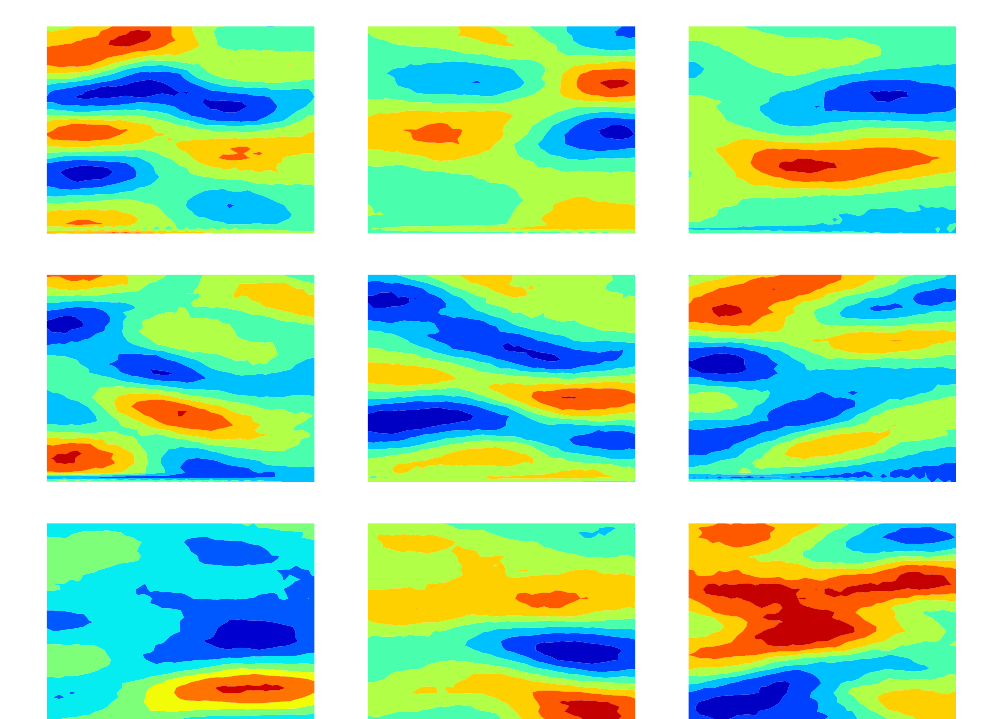}}
\subfloat[][]{\includegraphics[width=0.46\linewidth,trim={7cm 6cm 0 0},clip]{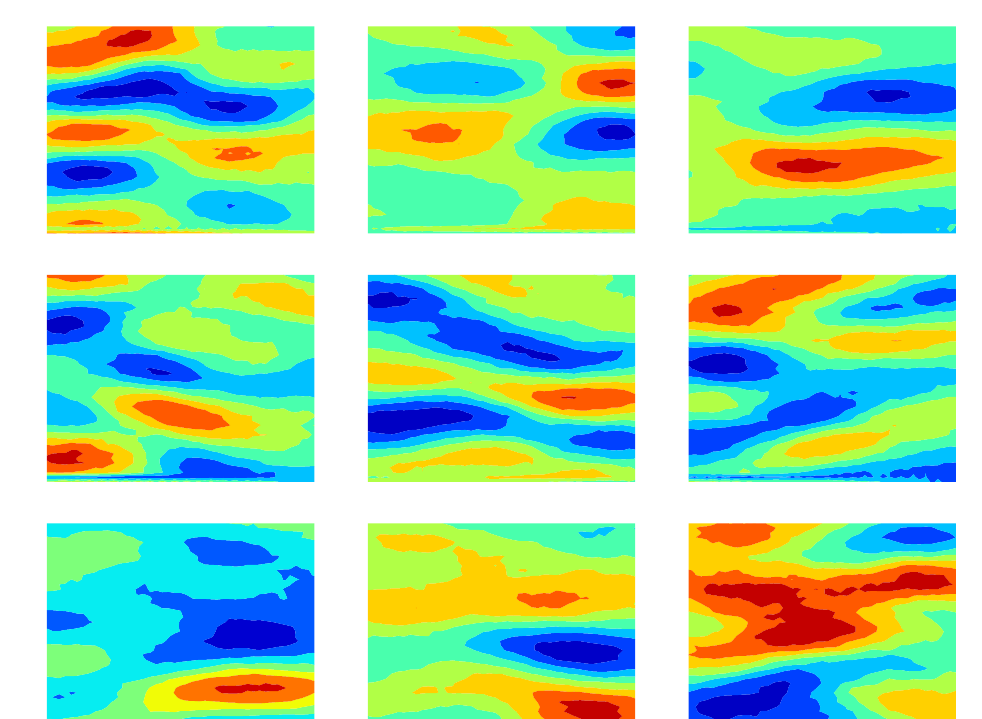}}\\
\subfloat[][]{\includegraphics[width=0.46\linewidth,trim={7cm 6cm 0 0},clip]{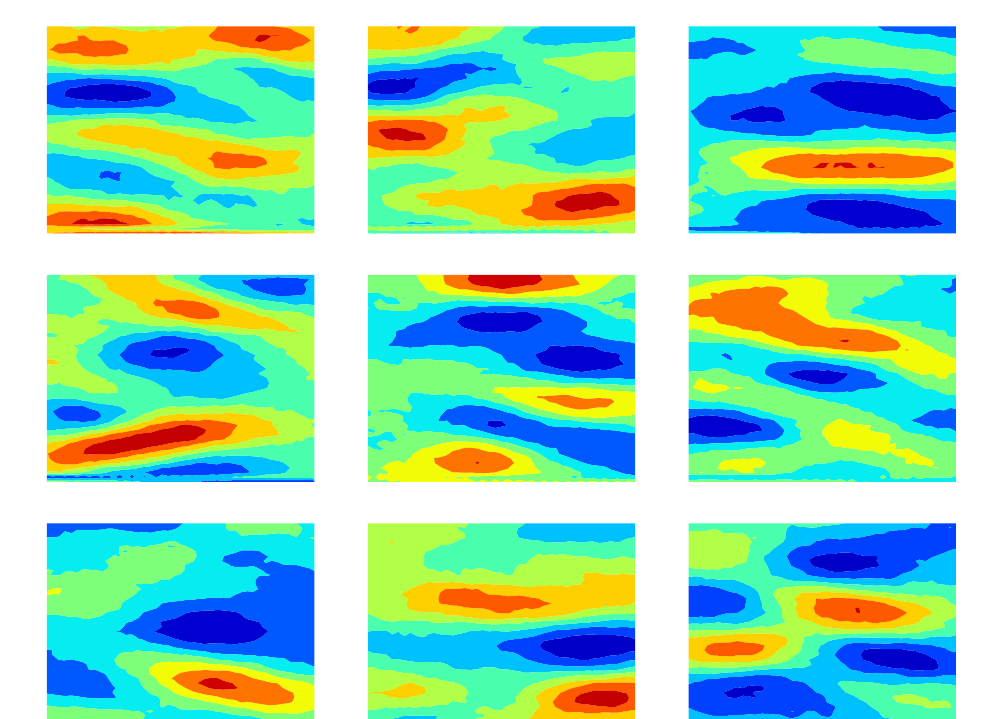}}
\subfloat[][]{\includegraphics[width=0.46\linewidth,trim={7cm 6cm 0 0},clip]{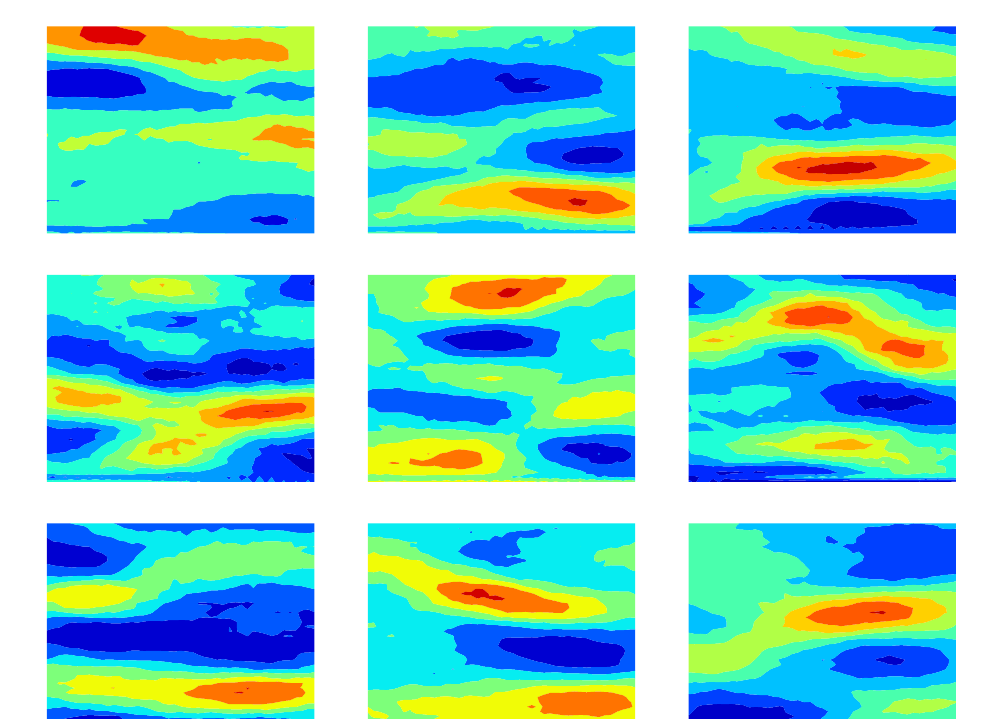}}\\
\subfloat[][]{\includegraphics[width=0.46\linewidth,trim={7cm 6cm 0 0},clip]{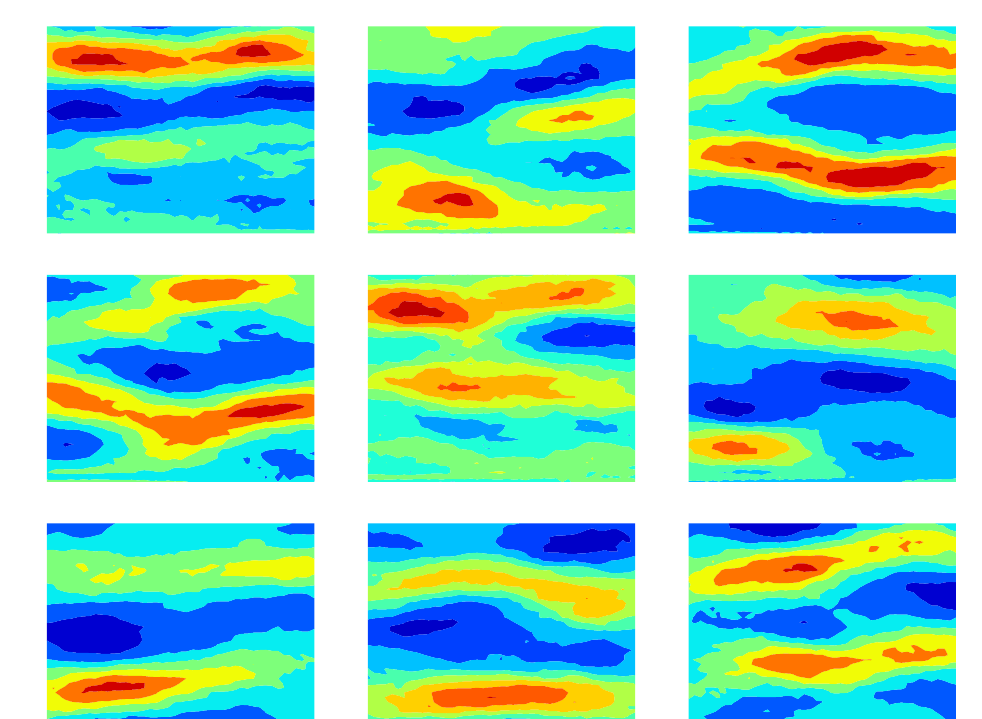}}
\subfloat[][]{\includegraphics[width=0.46\linewidth,trim={7cm 6cm 0 0},clip]{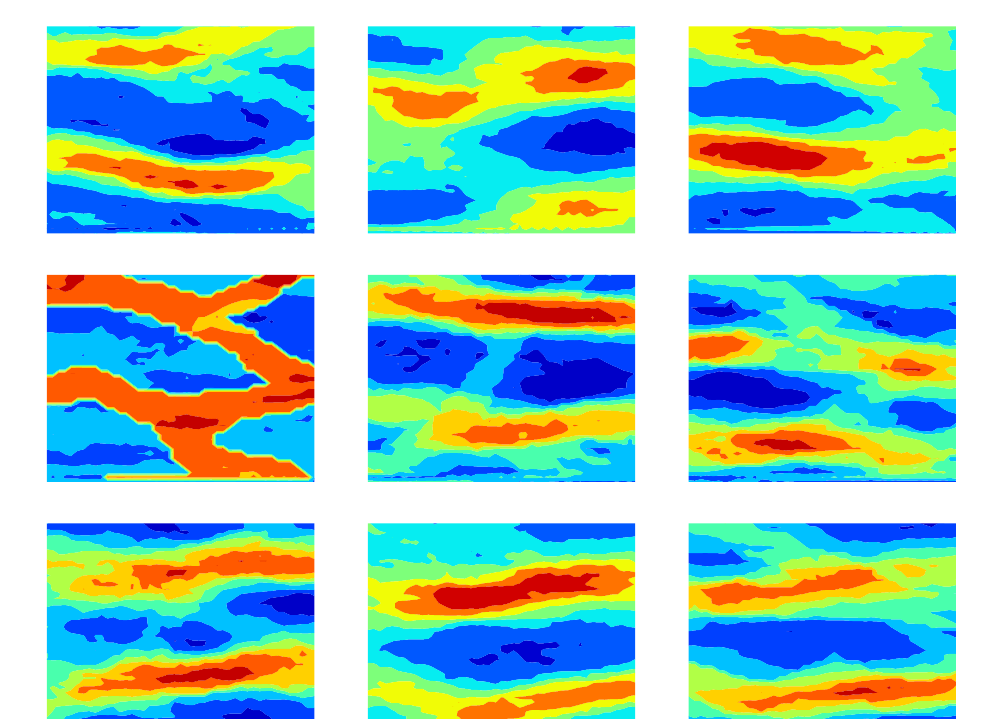}}
\caption{A few snapshots generated using mean perturbation in KPCA space with a)Gaussian b)linear c)quadratic d)cubic e)fourth order and f)fifth order kernels}\label{fig:pert}
\end{figure}
In contrast to linear PCA,  KPCA is performed in the feature space instead of the original space.
For the polynomial kernel $(\bfx\cdot \bfy )^d$, an input space of realization in $\bbR^{N_R}$ is mapped to a feature space of  dimension $N_F$ given by~\eqref{eq:NF}. Compared to the dimension of the original space $\bbR^{N_R}$, $N_F$ is very large with higher order polynomial kernels. For instance, in our channelized model, we have $N_R=10^3$ and for $d = 5$ this leads to $N_F\approx 10^{15}$, a very high-dimensional space which allows kernel PCA to explore and capture distinctive properties of the nonlinear data. Note here that the KPCA-feature space is still obtained by a low-dimensional eigendecomposition similar to PCA with the kernel trick. 

\begin{figure}[h]
\centering
\includegraphics[width=0.98\textwidth]{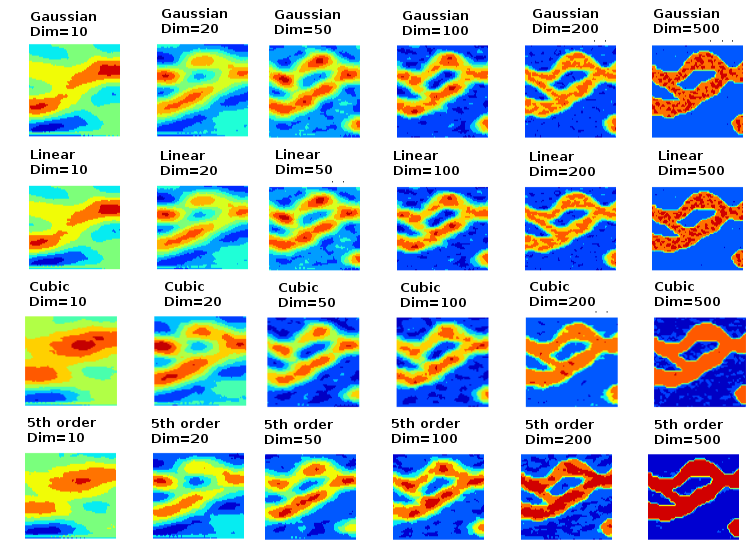}
\caption{KPCA with Gaussian, linear,  cubic and fifth order kernels in 10, 20, 50, 100, 200, 500, 1000 dimensions}\label{fig:check-kpca}
\end{figure}
Since our interest is to find inverse solutions in the original space, an additional pre-imaging step is required to transform the feature snapshots back into the original snapshots.  Unlike linear PCA, the solution to the pre-imaging is not unique and also suffers from instability. In order to choose the best kernel for our procedure, we test Gaussian, linear,  quadratic, cubic, 4th and 5th order polynomial kernels for their pre-imaging efficiency  using a few selected snapshots. Figure ~\ref{fig:check-kpca} depicts the results from this procedure for a pre-selected snapshot. It shows that higher order ($d$) polynomial kernels  lead to more efficient mapping. Also, we observed the computation of the pre-image became unstable for polynomial kernels order greater than five.
\begin{figure}[h]
\centering
\includegraphics[width=0.65\textwidth]{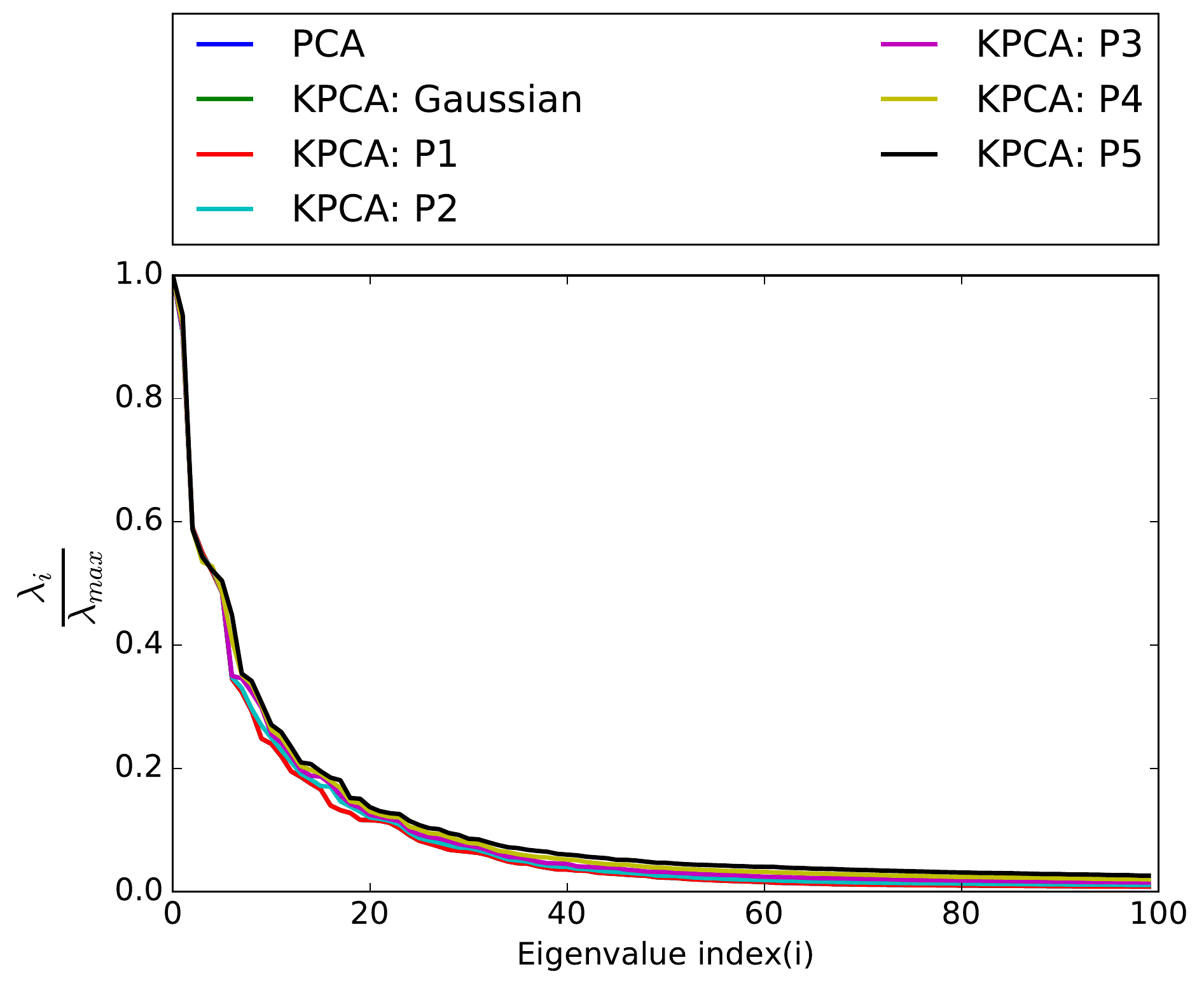}
\caption{Eigenvalue decay of the snapshots for different kernels}\label{fig:kpca2}
\end{figure}
Figure~\ref{fig:kpca2} shows the eigenvalue decay of the covariance matrix for Gaussian and polynomial kernels, showing that linear PCA and KPCA have similar eigen spectrums. Figure~\ref{fig:pert} displays a few snapshots generated using mean perturbation in KPCA space with Gaussian, linear, quadratic, cubic, fourth, and fifth order kernels. This demonstrates that as the order of the polynomial kernel increases, the mean perturbed data looks more like a channelized structure---i.e., higher order kernels are able to represent data more effectively.  Based on Figs.~\ref{fig:check-kpca}, \ref{fig:kpca2} and  \ref{fig:pert},   we select a polynomial kernel with order 5 and dimension 20 (about 75\% contribution).
\subsection{Efficiency of the PCE}
\begin{figure}[h]
\centering
\includegraphics[width=0.98\textwidth]{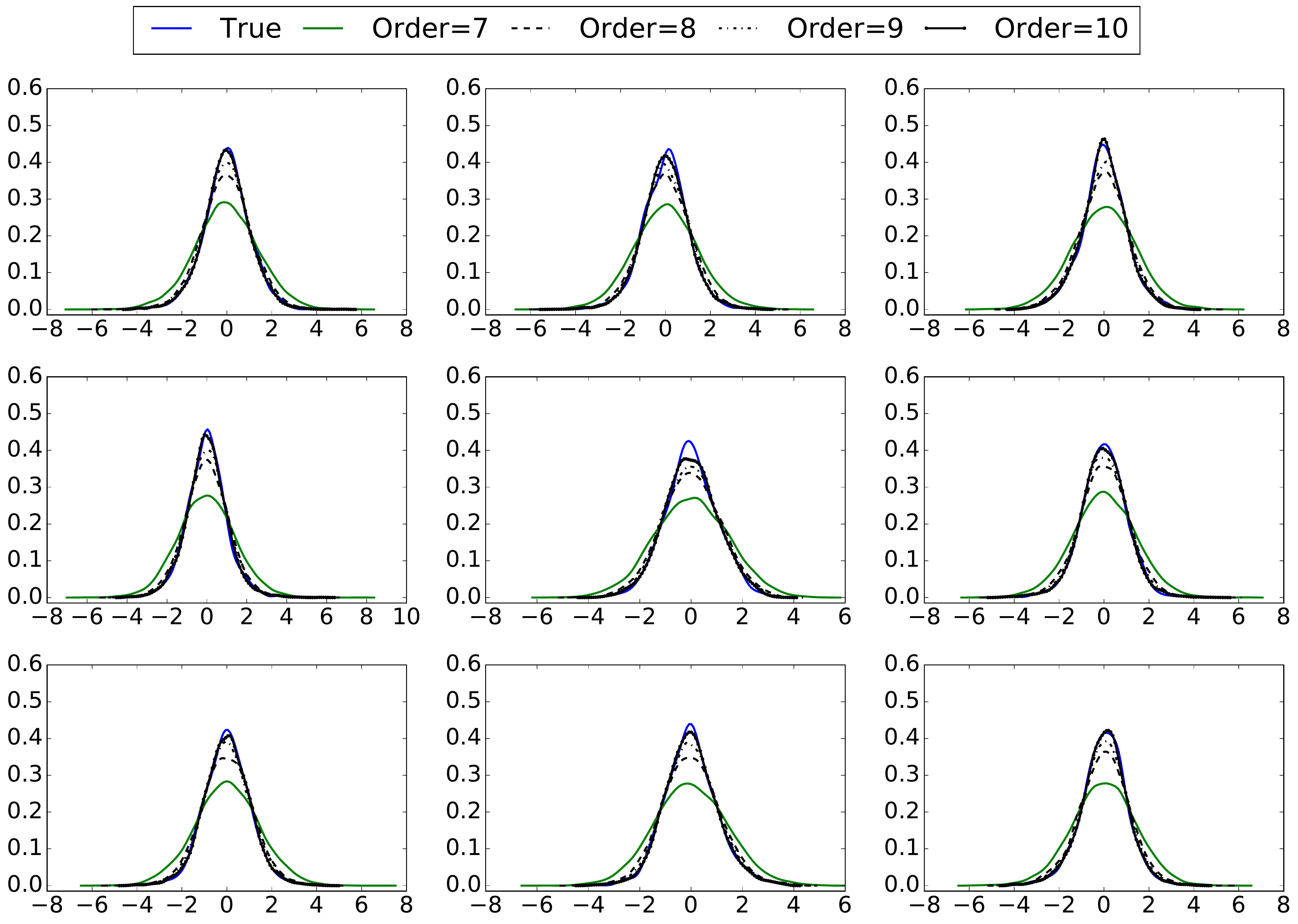}
\caption{Probability density function of a few $\xi^d$ obtained using true samples and from the samples of PCE with different orders}\label{fig:pce}
\end{figure}
Nonlinear mapping of the parameter space $\itPhi: \bbR^{N_R} \rightarrow \bbR^{N_F}, \: N_F \gg N_R$ and  solving~\eqref{discrete_xi} lead to 1000 discrete realizations of the $\bsxi^d$. In general, $\bsxi^d$ are
non-Gaussian, uncorrelated and dependent random variables. To generate these realizations in a computationally efficient way during the inversion procedure, assuming $\bsxi^d$ are independent similar to~\cite{ghanem2006construction,stefanou2009identification}, we construct multiple PCEs for $\bsxi^d$ using ICDF mapping.  Figure~\ref{fig:pce} depicts the probability density functions of a few selected $\xi^d$ constructed from the 1000 discrete realizations (true) and also samples obtained from the PCE with different orders. This figure demonstrates that, as the order of the PCE increases, PCE is able to capture the true distribution of the $\xi^d$. Based on this plot, the PCE with order 10 is used to map $\xi^d$ to the standard Gaussian variable $\eta$.


\subsection{Stochastic inversion using MHMCMC and Langevin MCMC}
The goal of our numerical demonstration is to recover the elastic parameters of the complex geological elasticity parameter field shown in  Fig.~\ref{fig:bay} (a). The ``ground truth" observations of displacements at the top, left and right boundary grid points are synthesized by running a forward simulator with aforementioned elasticity parameters. Due to sparsity of the measurements and low-dimensionality of the feature space we foresee that the posterior solution will converge to the lower dimensional version (Fig.~\ref{fig:bay} (b)) of the original snapshot. 
\begin{figure}[h]
\centering
\subfloat[][]{\includegraphics[width=0.46\linewidth]{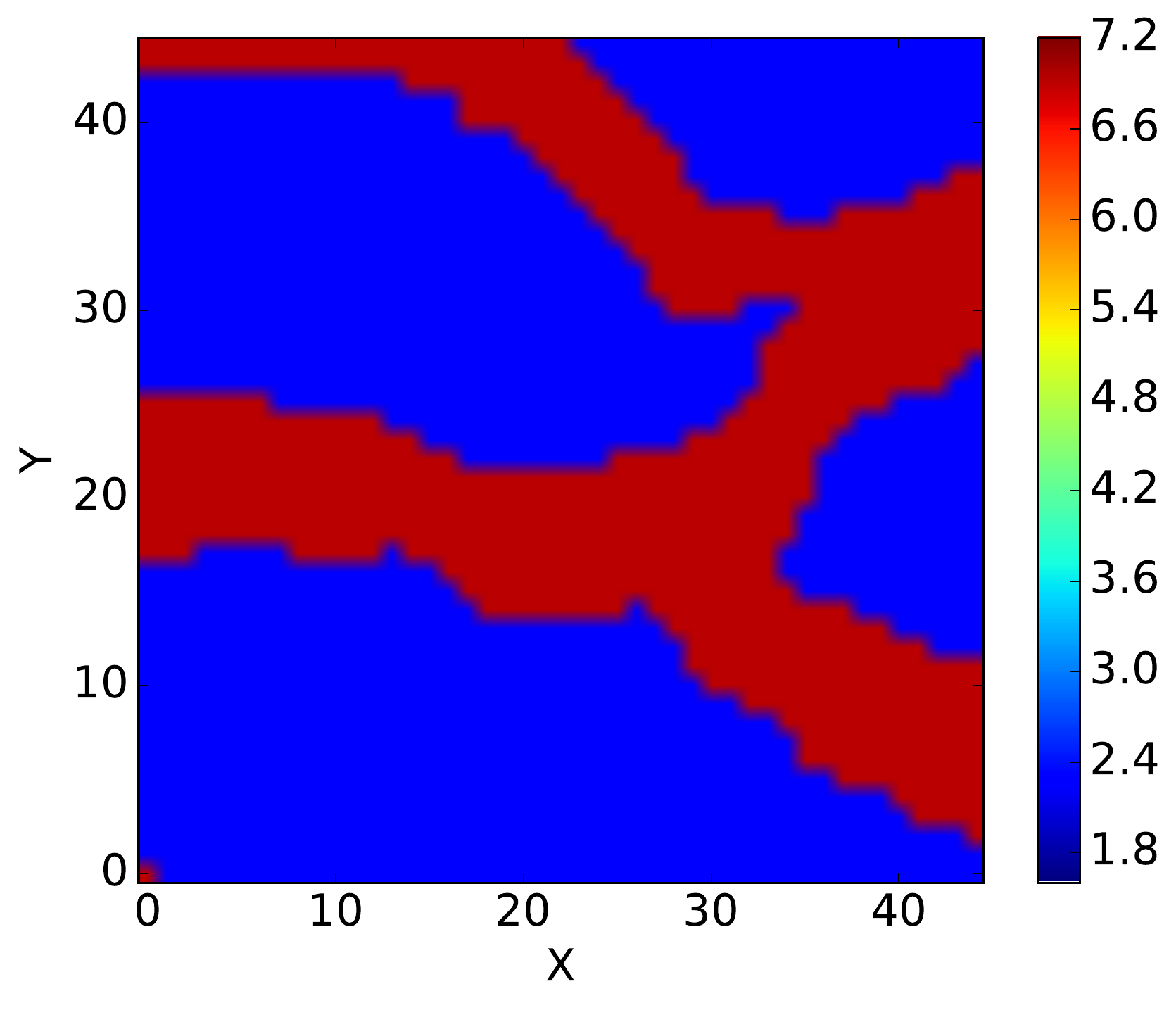}}
\subfloat[][]{\includegraphics[width=0.46\linewidth]{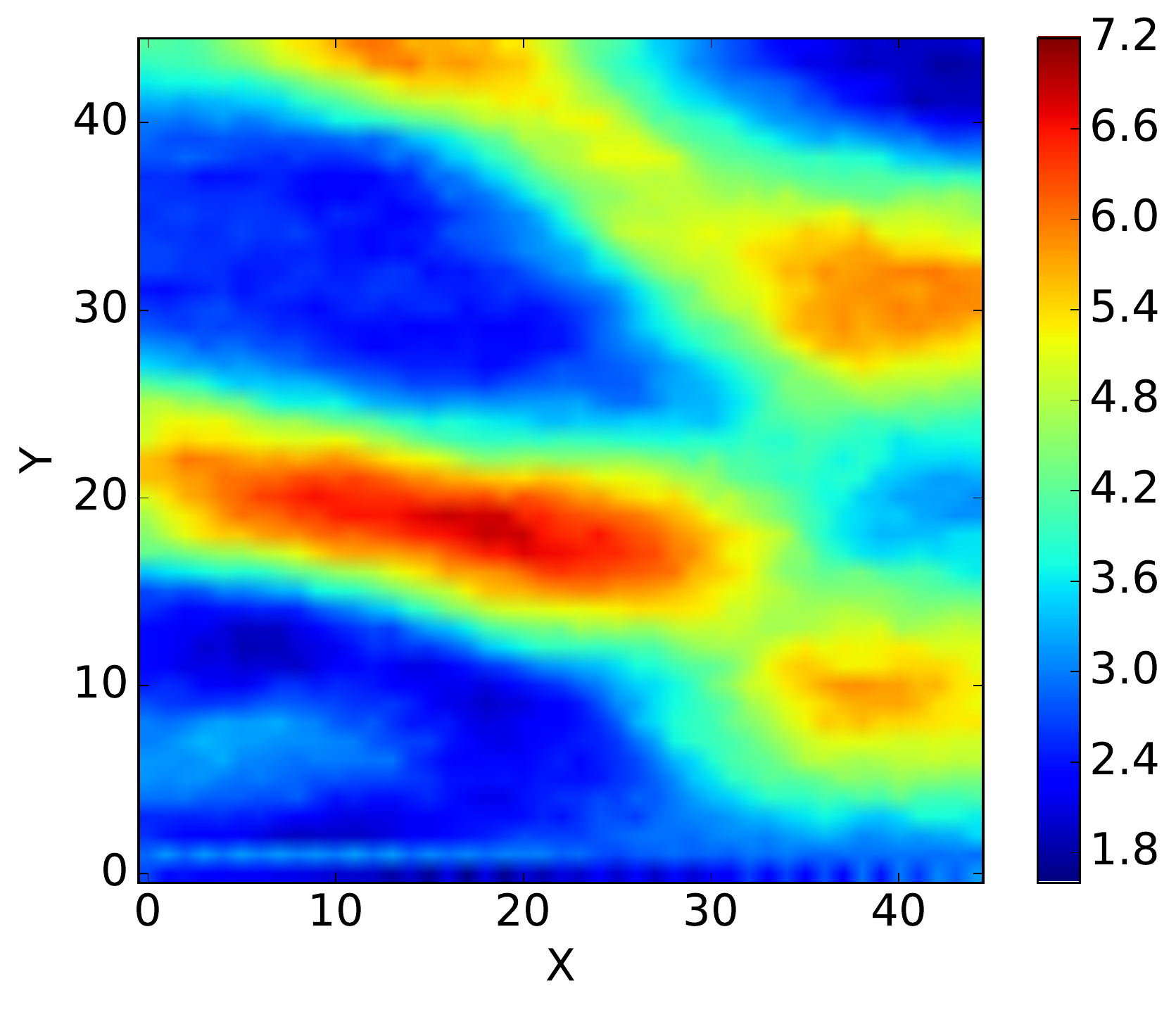}}\\
\subfloat[][]{\includegraphics[width=0.46\linewidth]{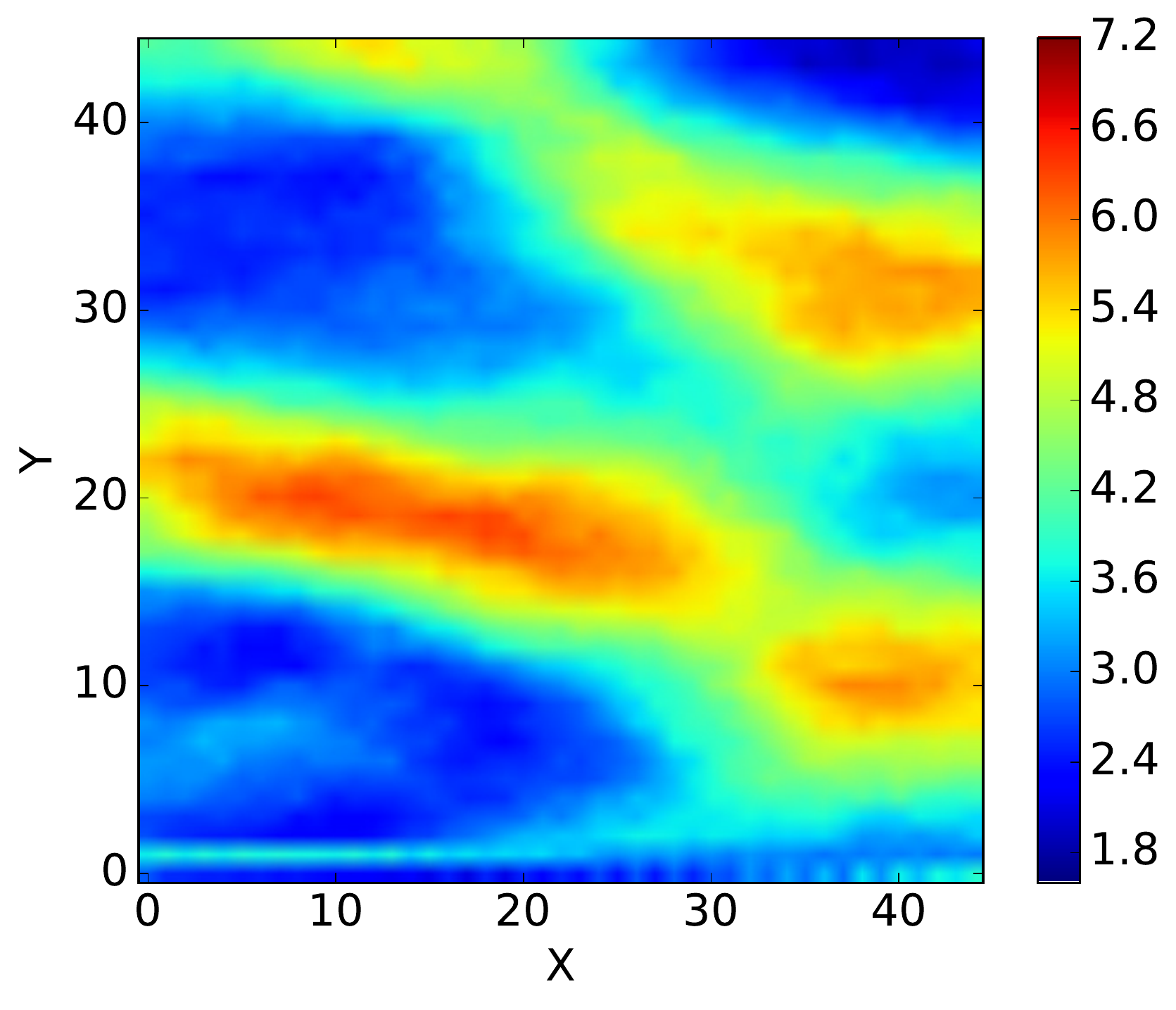}}
\subfloat[][]{\includegraphics[width=0.46\linewidth]{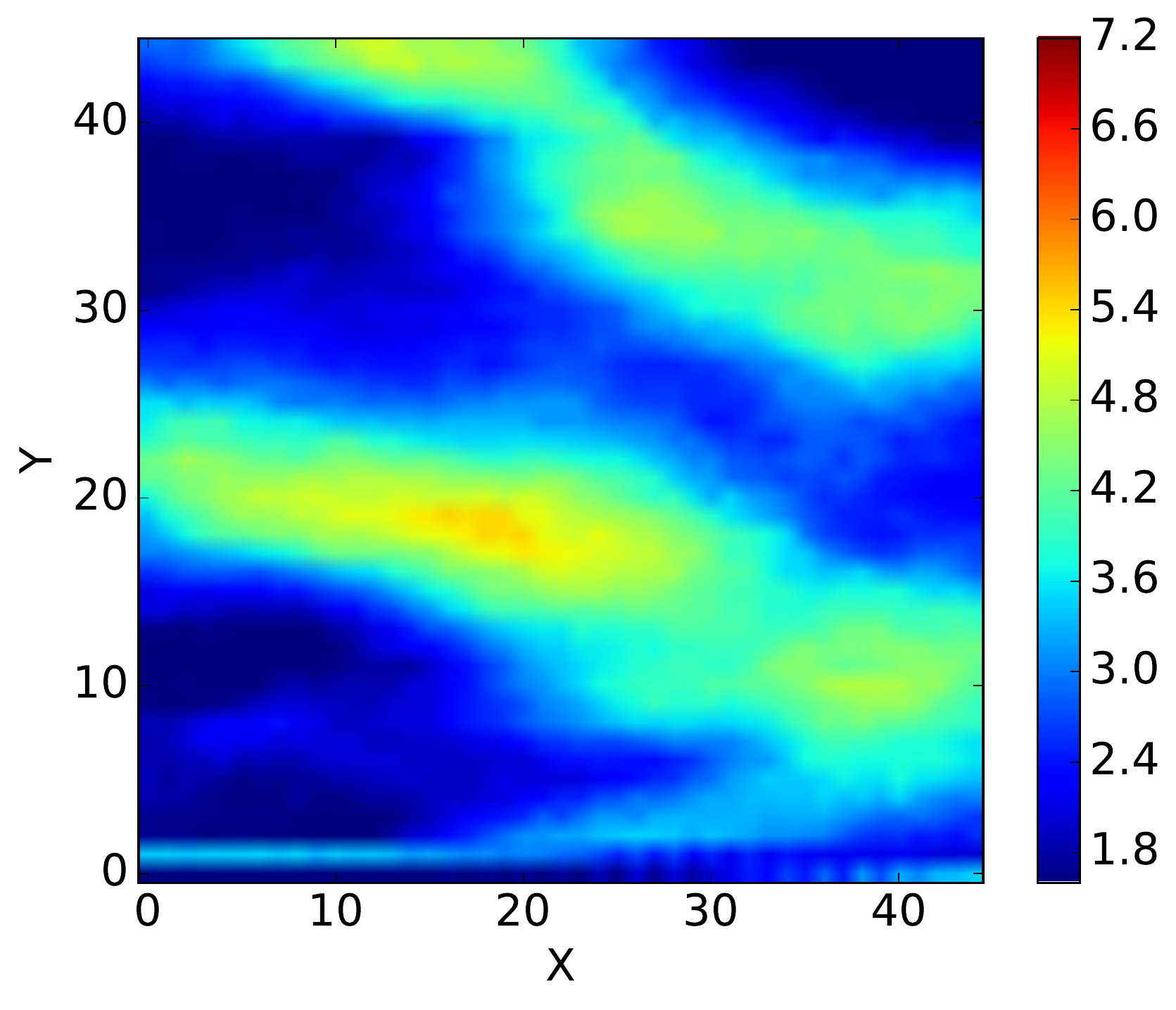}}\\
\subfloat[][]{\includegraphics[width=0.46\linewidth]{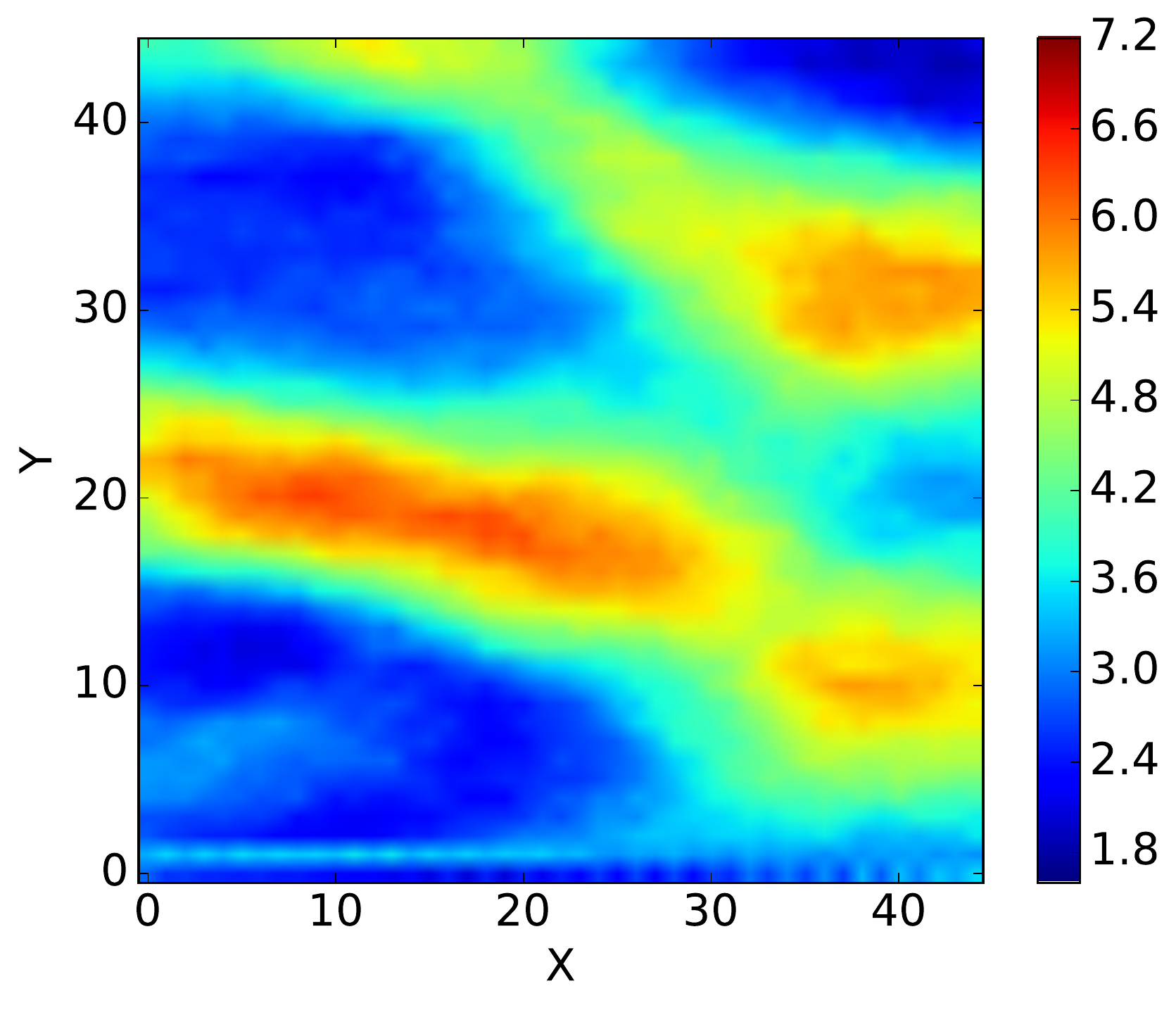}}
\subfloat[][]{\includegraphics[width=0.46\linewidth]{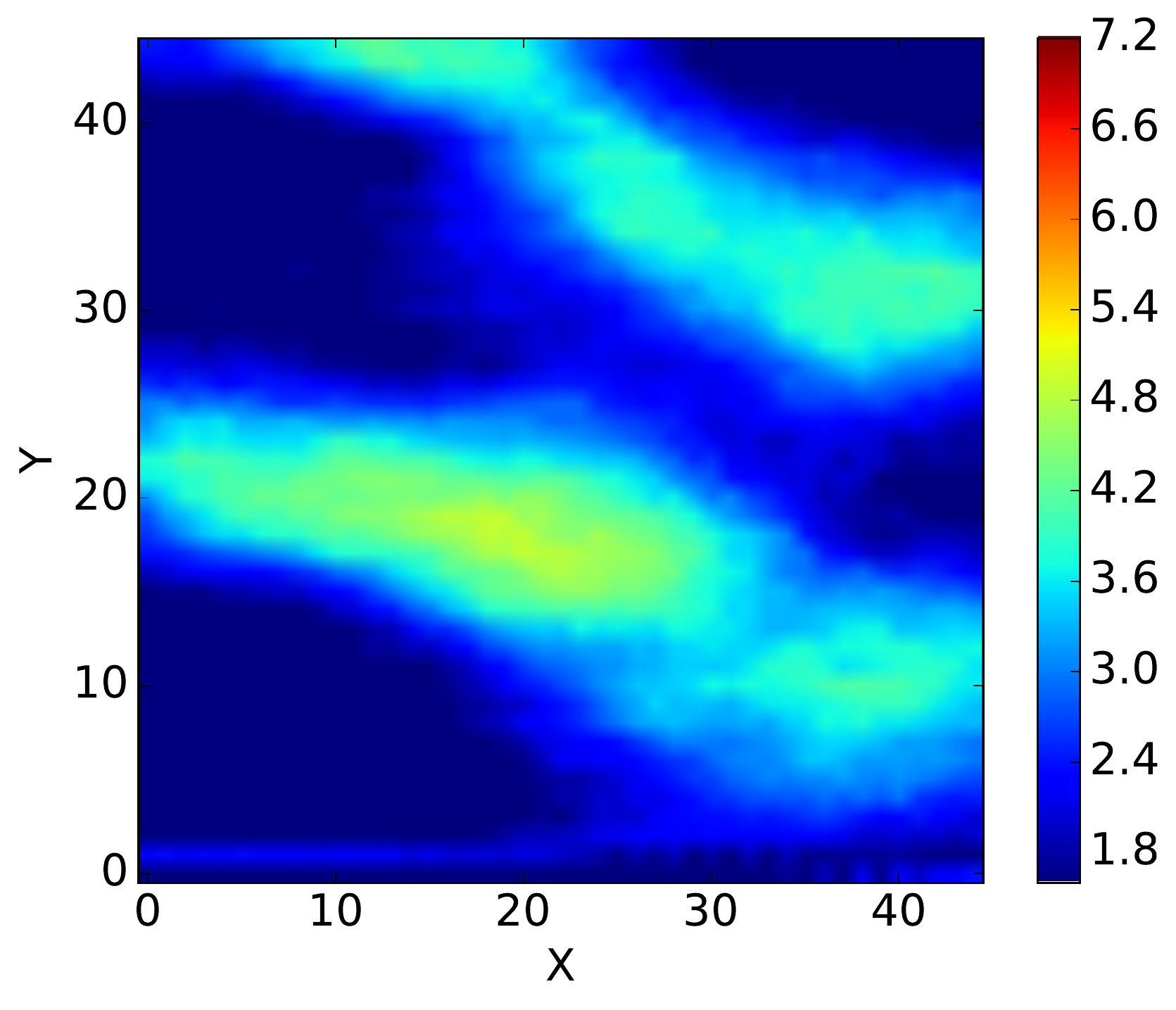}}
\caption{a) Original b) KPCA projected c) MHMCMC posterior mean d) MHMCMC posterior standard deviation e) Langevin MCMC posterior mean and f) Langevin MCMC posterior standard deviation snapshots }\label{fig:bay}
\end{figure}
\begin{figure}[h]
\centering
\subfloat[][]{\includegraphics[width=0.95\linewidth]{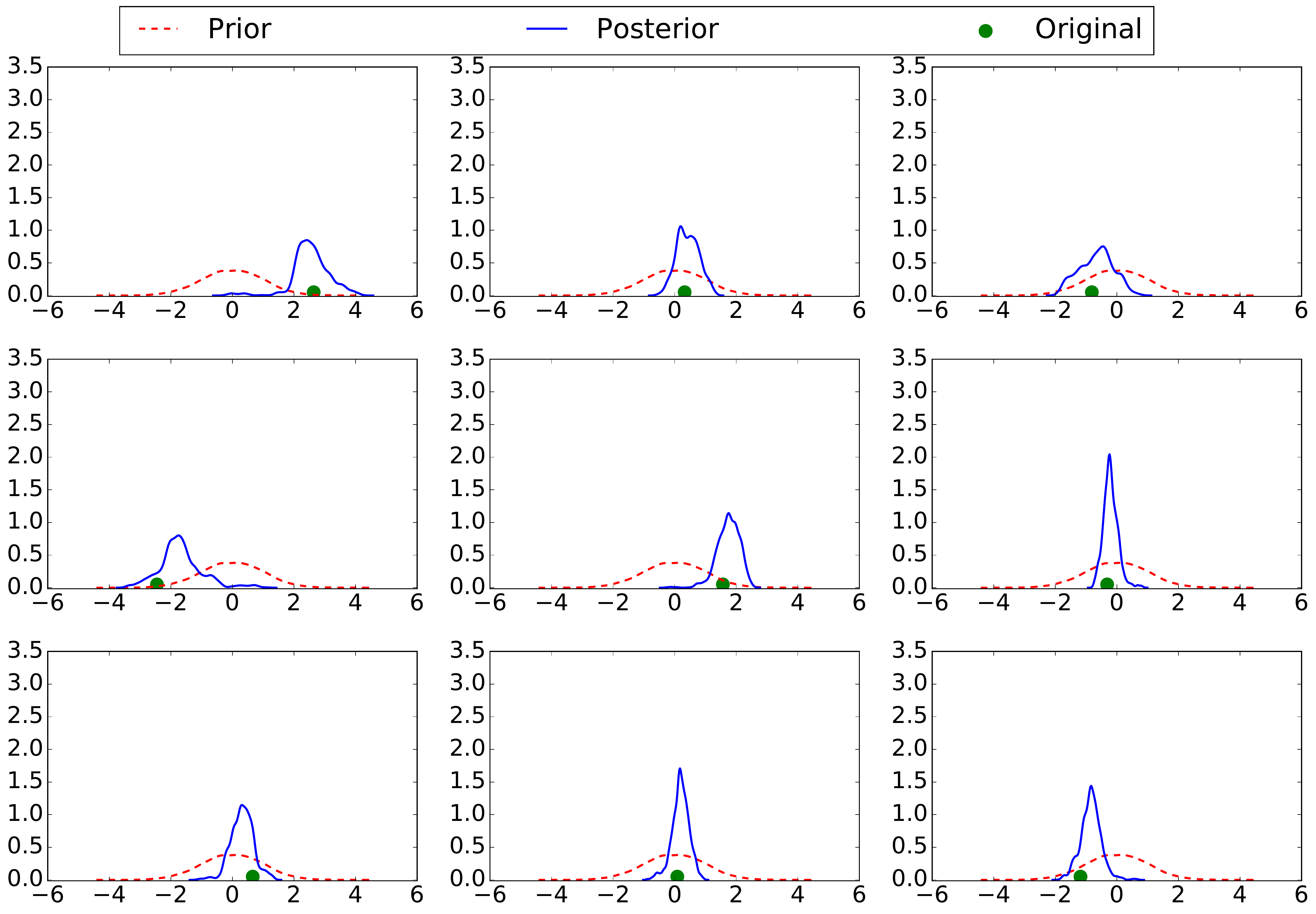}}\\
\subfloat[][]{\includegraphics[width=0.95\linewidth]{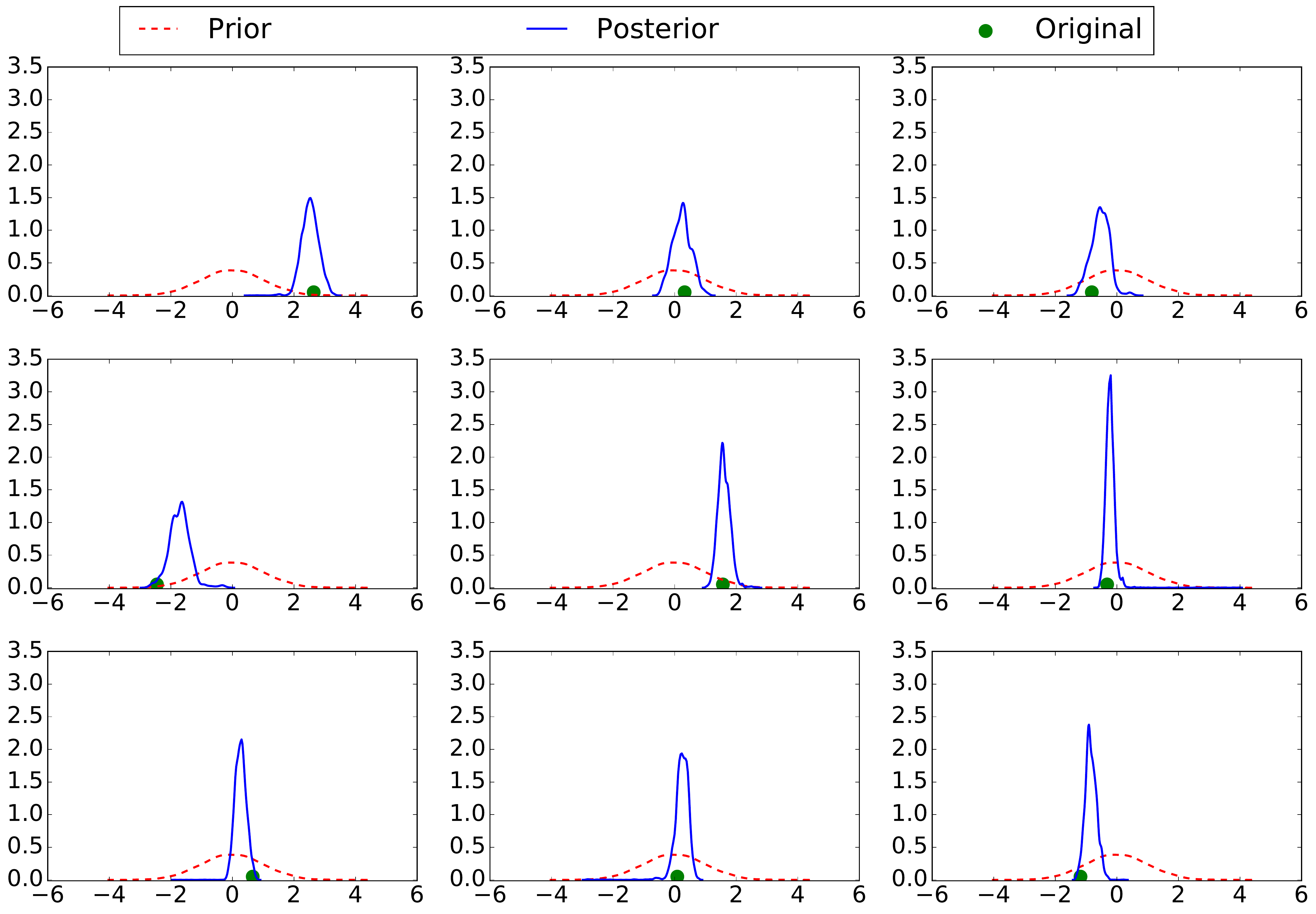}}
\caption{Prior and posterior probability density functions for a few $\eta$'s with original value for a) MHMCMC b) Langevin MCMC}\label{fig:pp}
\end{figure}

The samples of the posterior distribution  are obtained with LMCMC and random walk MHMCMC algorithms. 
Since the posterior exploration is carried out in $\bm\eta$ space, a multi-dimensional standard normal distribution servers as a prior distribution.  For MHMCMC, the proposal or sampling distribution is assumed to be Gaussian centered at current accepted sample with standard deviation of 0.1. The Langevin parameter $\tau$ is chosen as 0.08 based on trail and error and likelihood is scaled by 1000 to avoid floating point underflow errors. Figures~\ref{fig:bay} (c) and (d) show the  posterior mean and standard deviation  snapshots obtained using  MHMCMC. Similarly, Fig.~\ref{fig:bay} (e) and (f)  show the  posterior mean and standard deviation snapshots obtained using LMCMC. As envisioned before both MCMC and LHMCMC are able to recover the low-dimensional version of the original parameter field.  Figure~\ref{fig:pp} depicts the posterior distribution of the $\eta$ for the random walk MHMCMC and LMCMC. The detailed analysis of the posterior distribution is carried out in the next section. 
\begin{figure}[h]
\centering
\subfloat[][]{\includegraphics[width=0.98\linewidth]{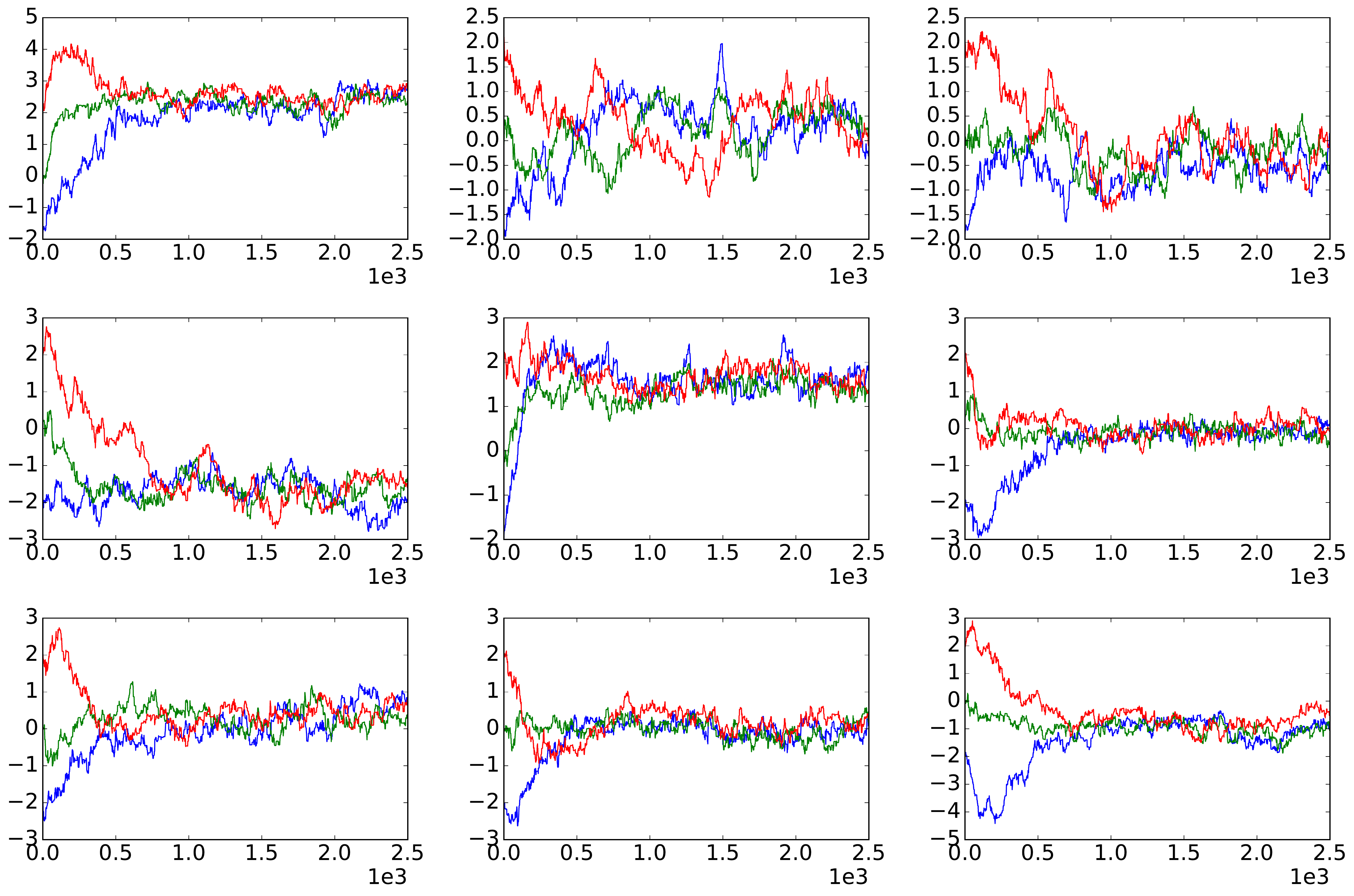}}\\
\subfloat[][]{\includegraphics[width=0.98\linewidth]{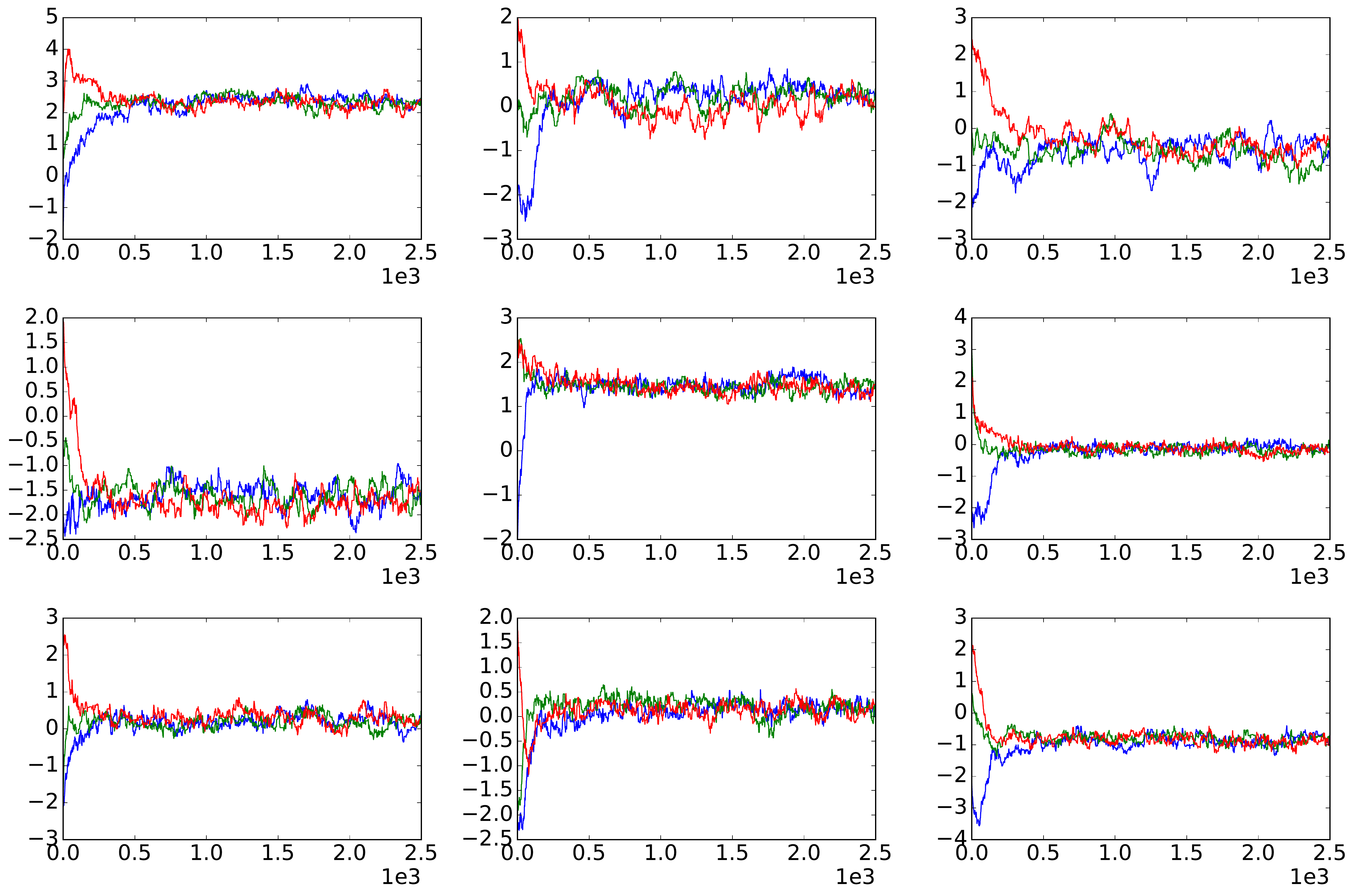}}
\caption{Posterior MCMC chains for a few $\eta$'s with staring at -2 (red), 0 (green) and 2(blue) for a)MCMC b)Langevin MCMC }\label{fig:chain}
\end{figure}
 Three MCMC chains with initial guess for $\eta$ as -2, 0 and 2 are used to check the global convergence of the MCMC algorithms.  Figure~\ref {fig:chain} shows the convergence of the MCMC chains for random walk MHMCMC and LMCMC. Chains start converging around the 100th and 500th sample for LMCMC and MHMCMC, respectively, i.e., gradient information assisted in substantially faster convergence.
 \section{Discussion}\label{sec:ds}
 \begin{figure}[h]
\centering
\includegraphics[width=0.99\linewidth]{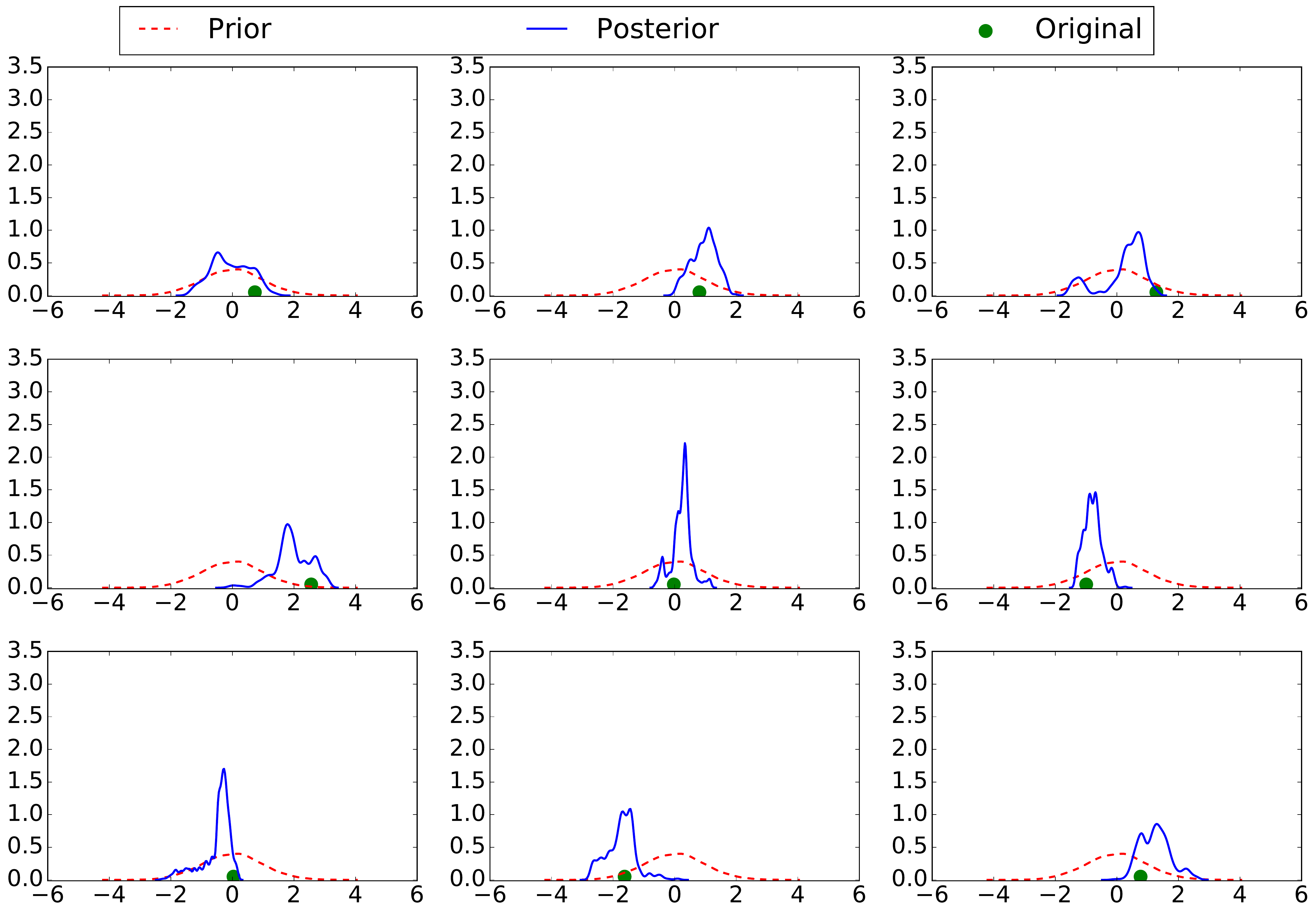}
\caption{Prior and posterior probability density functions for a few $\eta$'s with original value for obtained using PCA-based Langevin MCMC}\label{fig:ppca}
\end{figure}
 As shown in Figure~\ref{fig:kpca2}, dimension reduction via linear PCA and KPCA generally gives similar reduced orders  based on existing data points, but the reduced-order space they represent can be very different. Since the proposed method is based on LMCMC, which has computational complexity of $O(n^{1/3})$ compared to MHMCMC complexity of $O(n)$, its computational cost scales better.  To see the effect of KPCA on posterior sampling, we run a PCA-based LMCMC. The  KPCA-based LMCMC and PCA-based LMCMC have 33.66\% and 10.10\% acceptance rate, respectively.

As to why KPCA is more efficient than PCA, we propose the following explanation: the posterior probability density functions (PDFs) inverted by PCA-based MCMC (as seen in Figure \ref{fig:ppca}) are generally non-Gaussian and possibly multi-modal.  In contrast, those inverted using KPCA-based MCMC have near-normal distributions and are generally unimodal as a result of the embedded nonlinear mapping from the feature space to the parameter space.  Since it is generally more expensive (requires more iterations) to achieve convergence for non-standard PDFs with many peaks, the gradient-based MCMC, which approximates the posteriors by a local Gaussian, is expected to be more efficient. A second reason is that,
compared to linear PCA, the embedded manifold identified by the data-driven KPCA contains a more `concentrated' distribution of the underlying parameters that need to be inverted. Even though finding an optimal point (deterministic inversion) in the detected manifold may not be very distinguishable from stochastic inversion, the latter (stochastic inversion) performed in such a clustered manifold will be critical for achieving high performance and accuracy. Specifically, the neighborhood identified by linear PCA for any given channelized material parameter point may contain very few channelized structures, which can cause great difficulties for a high-dimensional random field inversions especially when considering stochastic inversions. Hence, the KPCA-based MCMC will demonstrate improved efficiency even without gradient information, thus making it useful even for the applications where the adjoint model cannot be derived easily.

As pointed out before, a relevant feature space identification for the problem considered here is analogous to a typical binary classification (channel vs no-channel) problem encountered in machine learning community.  
The discriminant function or boundary between two classes is linear in PCA---i.e., PCA detects a linear manifold in the original space. The KPCA or other kernel based methods such as diffusion maps, transforms data to a non-linear space with the kernel trick and detect a linear manifold in that space. Since the discriminant function deduced here is a linear function in terms of the weights, they detect a linear manifold in the non-linear space. In the future work, we will pursue feature space identification in the kernel space with 
non-linear or `curved' manifold learning using so-called deep autoencoders. Note here that KPCA and PCA can be described with autoencoder with a particular choice of activation function and decoding part of the deep network allows us to construct pre-imaging with a simple matrix-vector multiplication.     

\section{Conclusions}\label{sec:co}
We have presented an efficient stochastic inversion method in the framework of Bayesian inference based on an adjoint model, automatic differentiation, and Kernel PCA. The complexity of the MCMC is reduced through control reduction and efficient gradient computation. We demonstrate a practical way to characterize a full pdf assigned to each grid point of a discretized parametric random field based on prior knowledge or estimation of the random field and observational information of measurements data. To ensure the efficiency of the stochastic inversion, the control reduction is obtained  by performing Bayesian inference in a low-dimensional feature space captured via KPCA.  Different kernels such as Gaussian, first, second, third, fourth and fifth-order polynomials were tested and the kernel of KPCA is chosen based on snapshots obtained from the pre-imaging and mean perturbation. A PCE is devised for  economic sampling from the feature space. The proposed method uses a high-fidelity forward model and thus can avoid sub-optimal solutions computed using surrogate-based methods. A gradient based LMCMC method is adopted for posterior sampling using cheaply computed gradients with an adjoint model and automatic differentiation. The efficiency of the proposed method is demonstrated through a synthetic numerical example with the objective of recovering the subsurface elastic parameters of the complex geological channelized field. Gradient-free MCMC and LMCMC  were able to sample from the true posterior after 500 and 100 forward model runs, respectively. The KPCA-based MCMC results show a higher acceptance rate compared to the PCA-based MCMC, since the neighborhood identified by KPCA for any given channelized material parameter point contains more channelized structures. The method proposed has a generic nature and it can be adapted to other types of physics.  For example, in future work we will consider the application of the proposed framework to a large-scale seismic inversion problem. It should be pointed out that the KPCA is a linear manifold statistical learning on the kernel space constructed by the nonlinear transformation from the original space. In future work, we will pursue feature reduction for optimal control and stochastic inversion with a broader choice of unsupervised learning approaches---e.g. non-linear manifold statistical learning techniques such as diffusion maps and deep-learning based autoencoders.    
\section*{Acknowledgment}
This work was funded by the Laboratory Directed Research and Development (LDRD; 16-ERD-023) program
and conducted under the auspices of the U.S. Department of Energy by Lawrence Livermore National
Laboratory under contract DE-AC52-7NA27344.
\section*{References}
\bibliographystyle{elsarticle-num}
\bibliography{ReferenceBibTex}

\begin{thebibliography}{10}
\expandafter\ifx\csname url\endcsname\relax
  \def\url#1{\texttt{#1}}\fi
\expandafter\ifx\csname urlprefix\endcsname\relax\def\urlprefix{URL }\fi
\expandafter\ifx\csname href\endcsname\relax
  \def\href#1#2{#2} \def\path#1{#1}\fi

\bibitem{anandarajah2011computational}
A.~Anandarajah, Computational methods in elasticity and plasticity: solids and
  porous media, Springer Science \& Business Media, 2011.

\bibitem{kirtman2013near}
B.~Kirtman, S.~Power, A.~Adedoyin, G.~Boer, R.~Bojariu, I.~Camilloni,
  F.~Doblas-Reyes, A.~Fiore, M.~Kimoto, G.~Meehl, et~al., Near-term climate
  change: projections and predictability.

\bibitem{dagan2005subsurface}
G.~Dagan, S.~P. Neuman, Subsurface flow and transport: a stochastic approach,
  Cambridge University Press, 2005.

\bibitem{kennett2013seismic}
B.~Kennett, Seismic wave propagation in stratified media, ANU Press, 2013.

\bibitem{graves1996simulating}
R.~W. Graves, Simulating seismic wave propagation in 3d elastic media using
  staggered-grid finite differences, Bulletin of the Seismological Society of
  America 86~(4) (1996) 1091--1106.

\bibitem{kundur1994power}
P.~Kundur, N.~J. Balu, M.~G. Lauby, Power system stability and control, Vol.~7,
  McGraw-hill New York, 1994.

\bibitem{tarantola2005inverse}
A.~Tarantola, Inverse problem theory and methods for model parameter
  estimation, SIAM, 2005.

\bibitem{martin2012stochastic}
J.~Martin, L.~C. Wilcox, C.~Burstedde, O.~Ghattas, A stochastic newton mcmc
  method for large-scale statistical inverse problems with application to
  seismic inversion, SIAM Journal on Scientific Computing 34~(3) (2012)
  A1460--A1487.

\bibitem{green2001delayed}
P.~J. Green, A.~Mira, Delayed rejection in reversible jump metropolis-hastings,
  Biometrika (2001) 1035--1053.

\bibitem{mira2001ordering}
A.~Mira, Ordering and improving the performance of monte carlo markov chains,
  Statistical Science (2001) 340--350.

\bibitem{haario1999adaptive}
H.~Haario, E.~Saksman, J.~Tamminen, Adaptive proposal distribution for random
  walk metropolis algorithm, Computational Statistics 14~(3) (1999) 375--396.

\bibitem{haario2001adaptive}
H.~Haario, E.~Saksman, J.~Tamminen, An adaptive metropolis algorithm, Bernoulli
  (2001) 223--242.

\bibitem{tierney1999some}
L.~Tierney, A.~Mira, Some adaptive monte carlo methods for bayesian inference,
  Statistics in medicine 18~(1718) (1999) 2507--2515.

\bibitem{roberts1998optimal}
G.~O. Roberts, J.~S. Rosenthal, Optimal scaling of discrete approximations to
  langevin diffusions, Journal of the Royal Statistical Society: Series B
  (Statistical Methodology) 60~(1) (1998) 255--268.

\bibitem{haario2006dram}
H.~Haario, M.~Laine, A.~Mira, E.~Saksman, Dram: efficient adaptive mcmc,
  Statistics and computing 16~(4) (2006) 339--354.

\bibitem{parno2014transport}
M.~Parno, Y.~Marzouk, Transport map accelerated markov chain monte carlo, arXiv
  preprint arXiv:1412.5492.

\bibitem{ghanem2003stochastic}
R.~G. Ghanem, P.~D. Spanos, Stochastic finite elements: a spectral approach
  (2003).

\bibitem{marzouk2009stochastic}
Y.~Marzouk, D.~Xiu, A stochastic collocation approach to bayesian inference in
  inverse problems.

\bibitem{marzouk2007stochastic}
Y.~M. Marzouk, H.~N. Najm, L.~A. Rahn, Stochastic spectral methods for
  efficient bayesian solution of inverse problems, Journal of Computational
  Physics 224~(2) (2007) 560--586.

\bibitem{bliznyuk2012local}
N.~Bliznyuk, D.~Ruppert, C.~A. Shoemaker, Local derivative-free approximation
  of computationally expensive posterior densities, Journal of Computational
  and Graphical Statistics 21~(2) (2012) 476--495.

\bibitem{joseph2012bayesian}
V.~R. Joseph, Bayesian computation using design of experiments-based
  interpolation technique, Technometrics 54~(3) (2012) 209--225.

\bibitem{rasmussen2006gaussian}
C.~E. Rasmussen, Gaussian processes for machine learning.

\bibitem{funahashi1989approximate}
K.-I. Funahashi, On the approximate realization of continuous mappings by
  neural networks, Neural networks 2~(3) (1989) 183--192.

\bibitem{hornik1989multilayer}
K.~Hornik, M.~Stinchcombe, H.~White, Multilayer feedforward networks are
  universal approximators, Neural networks 2~(5) (1989) 359--366.

\bibitem{scholkopf1997kernel}
B.~Sch{\"o}lkopf, A.~Smola, K.-R. M{\"u}ller, Kernel principal component
  analysis, in: International Conference on Artificial Neural Networks,
  Springer, 1997, pp. 583--588.

\bibitem{Sarma2008}
P.~Sarma, L.~J. Durlofsky, K.~Aziz, Kernel principal component analysis for
  efficient, differentiable parameterization of multipoint geostatistics,
  Mathematical Geosciences 40~(1) (2008) 3--32.

\bibitem{Ma:2011:KPC:2016171.2016423}
X.~Ma, N.~Zabaras, Kernel principal component analysis for stochastic input
  model generation, J. Comput. Phys. 230~(19) (2011) 7311--7331.

\bibitem{strebelle2002conditional}
S.~Strebelle, Conditional simulation of complex geological structures using
  multiple-point statistics, Mathematical Geology 34~(1) (2002) 1--21.

\bibitem{marsden:1983}
J.~E. Marsden, T.~J.~R. Hughes, Mathematical Foundations of Elasticity, Dover
  Publications, New York, 1983.

\bibitem{MR2424078}
R.~A. Adams, J.~J.~F. Fournier, Sobolev spaces, 2nd Edition, Vol. 140 of Pure
  and Applied Mathematics (Amsterdam), Elsevier/Academic Press, Amsterdam,
  2003.

\bibitem{hughes:2000}
T.~J.~R. Hughes, The Finite Element Method: Linear Static and Dynamic Finite
  Element Analysis, Dover Publications, New York, 2000.

\bibitem{oberai2003solution}
A.~A. Oberai, N.~H. Gokhale, G.~R. Feij{\'o}o, Solution of inverse problems in
  elasticity imaging using the adjoint method, Inverse problems 19~(2) (2003)
  297.

\bibitem{Oberai:2004}
A.~A. Oberai, N.~H. Gokhale, M.~M. Doyley, J.~C. Bamber, Evaluation of the
  adjoint equation based algorithm for elasticity imaging, Physics in Medicine
  and Biology 49 (2004) 2955--2974.

\bibitem{courant1966methods}
R.~Courant, D.~Hilbert, Methods of mathematical physics, Vol.~1, CUP Archive,
  1966.

\bibitem{press2007numerical}
W.~H. Press, Numerical recipes 3rd edition: The art of scientific computing,
  Cambridge university press, 2007.

\bibitem{bishop2006pattern}
C.~M. Bishop, Pattern recognition, Machine Learning 128 (2006) 1--58.

\bibitem{cressie1990origins}
N.~Cressie, The origins of kriging, Mathematical geology 22~(3) (1990)
  239--252.

\bibitem{isaaks1989applied}
E.~H. Isaaks, et~al., Applied geostatistics, Tech. rep., Oxford University
  Press (1989).

\bibitem{matheron1963principles}
G.~Matheron, Principles of geostatistics, Economic geology 58~(8) (1963)
  1246--1266.

\bibitem{lim2003reservoir}
J.-S. Lim, Reservoir permeability determination using artificial neural
  network, J. Korean Soc. Geosyst. Eng 40 (2003) 232--238.

\bibitem{nikravesh2001past}
M.~Nikravesh, F.~Aminzadeh, Past, present and future intelligent reservoir
  characterization trends, Journal of Petroleum Science and Engineering 31~(2)
  (2001) 67--79.

\bibitem{nikravesh2003soft}
M.~Nikravesh, L.~A. Zadeh, F.~Aminzadeh, Soft computing and intelligent data
  analysis in oil exploration, Vol.~51, Elsevier, 2003.

\bibitem{ouenes2000practical}
A.~Ouenes, Practical application of fuzzy logic and neural networks to
  fractured reservoir characterization, Computers \& Geosciences 26~(8) (2000)
  953--962.

\bibitem{gholami2012prediction}
R.~Gholami, A.~Shahraki, M.~Jamali~Paghaleh, Prediction of hydrocarbon
  reservoirs permeability using support vector machine, Mathematical Problems
  in Engineering 2012.

\bibitem{Thimmisetty2017}
C.~A. Thimmisetty, R.~G. Ghanem, J.~A. White, X.~Chen, High-dimensional
  intrinsic interpolation using gaussian process regression and diffusion maps,
  Mathematical Geosciences.

\bibitem{Scholkopf:1998:NCA:295919.295960}
B.~Sch\"{o}lkopf, A.~Smola, K.-R. M\"{u}ller, Nonlinear component analysis as a
  kernel eigenvalue problem, Neural Comput. 10~(5) (1998) 1299--1319.

\bibitem{Schölkopf1997}
B.~Sch{\"o}lkopf, A.~Smola, K.-R. M{\"u}ller, Kernel principal component
  analysis, Springer Berlin Heidelberg, Berlin, Heidelberg, 1997, pp. 583--588.

\bibitem{MR2036084}
I.~T. Jolliffe, Principal component analysis, 2nd Edition, Springer Series in
  Statistics, Springer-Verlag, New York, 2002.

\bibitem{kwok2003pre}
J.~T. Kwok, I.~W. Tsang, The pre-image problem in kernel methods, in: ICML,
  2003, pp. 408--415.

\bibitem{bengio2004out}
Y.~Bengio, J.-f. Paiement, P.~Vincent, O.~Delalleau, N.~L. Roux, M.~Ouimet,
  Out-of-sample extensions for lle, isomap, mds, eigenmaps, and spectral
  clustering, in: Advances in neural information processing systems, 2004, pp.
  177--184.

\bibitem{arias2007connecting}
P.~Arias, G.~Randall, G.~Sapiro, Connecting the out-of-sample and pre-image
  problems in kernel methods, in: Computer Vision and Pattern Recognition,
  2007. CVPR'07. IEEE Conference on, IEEE, 2007, pp. 1--8.

\bibitem{honeine2009solving}
P.~Honeine, C.~Richard, Solving the pre-image problem in kernel machines: A
  direct method, in: Machine Learning for Signal Processing, 2009. MLSP 2009.
  IEEE International Workshop on, IEEE, 2009, pp. 1--6.

\bibitem{honeine2011preimage}
P.~Honeine, C.~Richard, Preimage problem in kernel-based machine learning, IEEE
  Signal Processing Magazine 28~(2) (2011) 77--88.

\bibitem{lebrun2009generalization}
R.~Lebrun, A.~Dutfoy, A generalization of the nataf transformation to
  distributions with elliptical copula, Probabilistic Engineering Mechanics
  24~(2) (2009) 172--178.

\bibitem{rosenblatt1952remarks}
M.~Rosenblatt, Remarks on a multivariate transformation, The annals of
  mathematical statistics 23~(3) (1952) 470--472.

\bibitem{ghanem2006construction}
R.~G. Ghanem, A.~Doostan, On the construction and analysis of stochastic
  models: characterization and propagation of the errors associated with
  limited data, Journal of Computational Physics 217~(1) (2006) 63--81.

\bibitem{stefanou2009identification}
G.~Stefanou, A.~Nouy, A.~Clement, Identification of random shapes from images
  through polynomial chaos expansion of random level set functions,
  International Journal for Numerical Methods in Engineering 79~(2) (2009)
  127--155.

\bibitem{wiener1938homogeneous}
N.~Wiener, The homogeneous chaos, American Journal of Mathematics 60~(4) (1938)
  897--936.

\bibitem{arnst2010identification}
M.~Arnst, R.~Ghanem, C.~Soize, Identification of bayesian posteriors for
  coefficients of chaos expansions, Journal of Computational Physics 229~(9)
  (2010) 3134--3154.

\bibitem{eldred2009comparison}
M.~Eldred, J.~Burkardt, Comparison of non-intrusive polynomial chaos and
  stochastic collocation methods for uncertainty quantification, AIAA paper
  976~(2009) (2009) 1--20.

\bibitem{NME:NME2546}
G.~Stefanou, A.~Nouy, A.~Clement, Identification of random shapes from images
  through polynomial chaos expansion of random level set functions,
  International Journal for Numerical Methods in Engineering 79~(2) (2009)
  127--155.

\bibitem{10.2307/2242455}
J.~L. Jin~Qin, Empirical likelihood and general estimating equations, The
  Annals of Statistics 22~(1) (1994) 300--325.

\bibitem{jones1990performance}
M.~Jones, The performance of kernel density functions in kernel distribution
  function estimation, Statistics \& Probability Letters 9~(2) (1990) 129--132.

\bibitem{giering1998recipes}
R.~Giering, T.~Kaminski, Recipes for adjoint code construction, ACM
  Transactions on Mathematical Software (TOMS) 24~(4) (1998) 437--474.

\bibitem{TapenadeRef13}
L.~Hasco{\"e}t, V.~Pascual, The {T}apenade {A}utomatic {D}ifferentiation tool:
  {P}rinciples, {M}odel, and {S}pecification, {ACM} {T}ransactions {O}n
  {M}athematical {S}oftware 39~(3).

\bibitem{thimmisetty2015multiscale}
C.~Thimmisetty, A.~Khodabakhshnejad, N.~Jabbari, F.~Aminzadeh, R.~Ghanem,
  K.~Rose, J.~Bauer, C.~Disenhof, Multiscale stochastic representation in
  high-dimensional data using gaussian processes with implicit diffusion
  metrics, in: Dynamic Data-Driven Environmental Systems Science, Springer
  International Publishing, 2015, pp. 157--166.

\end{thebibliography}
\end{document}